\documentclass[letterpaper,twoside,british,english,american,review]{elsarticle}
\usepackage[T1]{fontenc}
\usepackage{xcolor}
\usepackage{pdfcolmk}
\usepackage{array}
\usepackage{float}
\usepackage{rotfloat}
\usepackage{booktabs}
\usepackage{textcomp}
\usepackage{multirow}
\usepackage{amstext}
\usepackage{amssymb}
\usepackage{graphicx}
\PassOptionsToPackage{normalem}{ulem}
\usepackage{ulem}

\makeatletter

\pdfpageheight\paperheight
\pdfpagewidth\paperwidth

\providecommand{\tabularnewline}{\\}
\providecolor{lyxadded}{rgb}{0,0,1}
\providecolor{lyxdeleted}{rgb}{1,0,0}

\journal{Example: Energy and Buildings}

\@ifundefined{showcaptionsetup}{}{%
 \PassOptionsToPackage{caption=false}{subfig}}
\usepackage{subfig}
\makeatother

\usepackage{babel}
\begin{document}

\title{Application of sensitivity analysis in building energy simulations:
combining first- and second-order elementary effects methods.}

\author[rvt]{D.~Garcia Sanchez}

\author[rvt]{B.~Lacarrière\fnref{fn2}\foreignlanguage{english}{\corref{cor1}}}

\ead{bruno.lacarriere@mines-nantes.fr, FAX +33 (0) 251 85 8299 }

\author[focal]{M.~Musy}

\author[rvt]{B.~Bourges}

\address[rvt]{Ecole des Mines de Nantes, UMR GEPEA - FR IRSTV, 4 rue Alfred Kastler,
44300, Nantes, France. }

\address[focal]{Ecole Nationale Supérieure d\textquoteright{}Architecture de Nantes,
UMR CERMA - FR IRSTV, 6 Quai F. Mitterrand 44000 Nantes, France. }
\begin{abstract}
Sensitivity analysis plays an important role in the understanding
of complex models. It helps to identify the influence of input parameters
in relation to the outputs. It can also be a tool to understand the
behavior of the model and can then facilitate its development stage.
This study aims to analyze and illustrate the potential usefulness
of combining first and second-order sensitivity analysis, applied
to a building energy model (ESP-r). Through the example of an apartment
building, a sensitivity analysis is performed using the method of
elementary effects (also known as the Morris method), including an
analysis of the interactions between the input parameters (second-order
analysis). The usefulness of higher-order analysis is highlighted
to support the results of the first-order analysis better. Several
aspects are tackled to implement the multi-order sensitivity analysis
efficiently: interval size of the variables, the management of non-linearity
and the usefulness of various outputs.\end{abstract}
\begin{keyword}
Energy demand \sep Buildings \sep Sensitivity analysis \sep Morris
method \sep Elementary effects \sep Building thermal simulation
\end{keyword}
\maketitle

\section{Introduction}

Energy consumption related to the building sector is recognized as
a major part of the total energy consumption worldwide (37\% of the
final energy consumption in the EU in 2004) \citep{PerezLombard2008394}
and consequently a significant source of greenhouse gas emissions
\citep{Salat2009}. The growth in population, building services and
comfort levels guarantees that this tendency will continue in the
forthcoming years. Many tools have been developed to model the energy
consumption in buildings (EnergyPlus, TRNSYS, ESP-r), particularly
for end uses such as space heating and cooling, ventilation and lighting.
In most cases, such models take into account coupling between phenomena
(e.g. interactions between occupancy, micro-climate, envelope, and
HVAC...) through coupling of different specialized sub-models and
by using a large number of diverse input variables.

Sensitivity analysis can help in understanding the relative influence
of input parameters on the output \citep{Lomas1992}. In the field
of building energy models, combining sensitivity analysis and simulations
tools can be useful as it helps to rank the input parameters (or family
of parameters) and then to select the most appropriate to be considered,
depending on the objective of the modeling. For example, this is particularly
interesting when the modeling objective is related to the building
design (e.g. sketch stage of the design, modeling retrofitting scenarios
according to the only available input data) or when it is to define
archetypes. Another application is in the development stage of the
tools, and more precisely the definition of possible simplification
of models or in the validation of assumptions in the selection of
input parameters that must be considered. In these cases, and depending
on the objectives of the tool developed, some sub-models and their
corresponding input data may become secondary. A solution consists
of using a detailed model in the upstream stage, combined with a sensitivity
analysis in order to rank the set of parameters and identify the coupling
between them. Then, the selection of the most important variable helps
to define the structure of the simplified model.

In this study, we propose combining the implementation of ESP-r \citep{Strachan2008601}
with two sensitivity analysis techniques: the Morris method \citep{Morris1991}
and an extension of this methodology for the analysis of interactions
between the parameters \citep{Campolongo19991}. In section 2, sensitivity
analysis methods are quickly reviewed. Then, the elementary effects
method and its second-order variant are described, together with the
apartment building test case (section 3). Finally, results are presented
and discussed for the two methods used (section 4).

\section{Background}

\subsection{Sensitivity analysis : current approaches}

Sensitivity analysis methods have been studied by many authors in
the past decades as they have demonstrated their strength in many
sectors. Throughout this period new methods and improvements have
been developed, offering different solutions depending on the objective.
Hamby \citep{Hamby1994} proposed an inventory of techniques for parameter
sensitivity analysis which he divided into three different categories: 

- Sensitivity analysis methods assessing the influence of individual
parameters. These include \emph{Differential Sensitivity Analysis,
One-at-a time sensitivity measures, Factorial Design, Sensitivity
Index, Importance Factors, and Subjective Sensitivity Analysis.}

- Parameter sensitivity analysis utilizing random sampling methods
(simple random sampling, Monte Carlo, Latin Hypercube). In this group
are listed the methods: \emph{Scatter plots, Importance Index, \textquoteleft{}Relative
Deviation\textquoteright{}, \textquoteleft{}Relative Deviation Ratio\textquoteright{},
Pearson\textquoteright{}s $r$, Rank Transformation, Spearman\textquoteright{}s
$\rho$, Partial Correlation Coefficient, Regression, and Standardized
Regression techniques. }

- Sensitivity tests involving segmented input distributions: \emph{the
Smirnov test, the Cramer Von-Mises test, the Mann-Whitney test, and
the squared-ranked test.} 

The author then applied these different methods to a case study related
to the nuclear industry, in order to compare them in terms of reliability,
computational requirements and ease of implementation \citep{Hamby1995}.
The study identified the One-at-a time method as being the simplest
but pointed out that it becomes time-intensive with large numbers
of parameters. Saltelli et al. \citep{saltelli2008} also describe
the different sensitivity analysis techniques. For these authors,
techniques can be divided between global and local methods. Local
methods are commonly based on the estimation of partial derivatives
in order to obtain a qualitative analysis of the importance of each
factor on the output response for a limited subset and particular
values of the input variables. Global methods vary all the parameters
and try to obtain information for a subset of input variables in a
wider domain. Global methods can also be divided into quantitative
and qualitative techniques. Santner et al. \citep{Santer2003} developed
the parallel between the physical experiments and the concept of computational
experiments as it is understood in this study. In particular, these
authors described the added value of sensitivity analysis in such
an experiment. This was taken up by Saltelli et al. \citep{saltelli2008}
who presented the One-at-a time sampling for sensitivity analysis
for multiple parameters.

\subsection{Principles of the elementary effects method}

The Morris method is derived from OAT (One-factor-at-a-time) screening
methods to identify the subset of the main important input factors
among a large number of $k$ input parameters in a model. This method
characterizes the sensitivity of a model with respect to its input
variables through the concept of \textit{elementary effects}, which
are approximations of the first order partial derivatives of the model
\citep{Morris1991}. These elementary effects are estimated at various
sampled points, randomly selected on a $p$-values regular grid, defining
a relevant design of computational experiments. The average and standard
deviations of elementary effects enable negligible and influencing
variables to be sorted and linear and non-linear influences to be
distinguished. In some respect, this method can be considered intermediate
between a local sensitivity analysis and global quantitative methods
described above. It is a general approach (model-independent), which
achieves a good compromise between accuracy and efficiency. Applications
can be found in a number of fields including Environmental Modeling
and Agriculture \citep{Richter2010127}, Biophysics \citep{Cooling20073421}
and Nuclear Engineering \citep{Nuclear1999}. However, in spite of
its advantages, its applications still remain limited.

Other methods, such as variance-based sensitivity indices (VBM), have
been proposed \citep{Saltelli2010259}. Although they generally provide
better information to distinguish non-linearities and interactions
\citep{Kucherenko20091135}, the computational cost is much higher:
a variance-based analysis for a 12-input parameter model requires
at least 14,000 runs of the model, about one hundred times the cost
of a first-order Morris analysis (and still ten times more than a
second-order Morris analysis).

\subsection{Experience with the elementary effects methods for building thermal
simulation}

Some studies have tested the advantages of the Morris method applied
to building energy simulations. Breesch and Janssens \citep{Breesch2004}
implemented it to identify the most important parameters that cause
uncertainty in the predicted performances of natural night ventilation.
They used a two-zone model in the thermal simulation tool TRNSYS coupled
with an infiltration model COMIS. This analysis revealed that the
internal heat gains, local outdoor temperature and the diurnal internal
convective heat transfer coefficient were the parameters with the
greatest impact on thermal comfort. Brohus et al. \citep{Brohus2009}
applied the methodology to reduce a set of 75 parameters used to obtain
an accurate output energy consumption distribution. The Morris method
was also used as a first indication of correlation or non-linear effects
between the parameters. Finally, this method was compared with the
Fourier Amplitude Sensitivity Testing method (FAST). The results of
both analysis helped in evaluating a safety factor for the annual
energy consumption at the design level. De Witt \citep{Witt1997}
compared the Morris method with the sequential bifurcation technique
using a mono-zone office of 81 parameters as a model. Both techniques
found the same set of important parameters (12) which explained 94\%
of the variability of the model output defined as the number of hours
of overheating. Corrado and Mechri \citep{Corrado2009} analyzed the
heating and cooling needs of a two-storey single-family house in Turin
with the Morris method to calculate the uncertainties in energy rating.
The sensitivity analysis showed that only 5 of 129 factors were responsible
for most of these uncertainties: the indoor temperature, the air change
rate, the number of occupants, the metabolism rate and the equipment
heat gains. Heiselberg et al. \citep{Heiselberg2009} identified the
most important design parameters in relation to a building's performance
with a focus on the optimization of sustainable buildings. They found
that the mechanical ventilation rate in winter and lighting control
were the most influential parameters in an office building of 7 floors.

The extension of the Morris method for second- and upper-order analysis
has still not been applied in the area of building energy simulations
despite its advantages stated in other analysis and its low computational
cost compared to more sophisticated techniques like variance-based
and FAST methods. 

\selectlanguage{english}%

\section{Methodology}

\selectlanguage{american}%

\subsection{The elementary effects method}

The building thermal model can be represented by a function $y(x)$
where $y$ is the output variable of interest (scalar) and $x$ is
a vector of real input variables with $k$ coordinates, each input
variable being defined within the range of a continuous interval.
Input variables are transformed into reduced dimensionless variables
in the interval $\left(0;1\right)$ as $x_{i}^{'}=\frac{x_{i}-x_{min}}{x_{max}-x_{min}}$.

$x_{min}$ and $x_{max}$ are the minimum and maximum of the input
variable $x_{i}$, respectively. The domain of the vector $x$ is
then a hypercube $H^{k}$ with unit length, a subset of $\mathbb{R}^{k}$.
For each reduced input variable, only discretized values are considered,
using a $p_{i}$ values regular grid (with step $\frac{1}{p_{i}-1}$):
$0$; $\frac{1}{p_{i}-1}$;$\frac{2}{p_{i}-1}$;...; $1$. Although
a single grid value is generally used for all the variables, it is
possible to use a specific one for each input variable $x_{i}$. This
enables qualitative input variables with two levels to be incorporated,
represented in the Morris design by a 2-value grid, while keeping
a more precise grid for continuous input variables.

A simulation trajectory is defined as a sequence of $\left(k+1\right)$
points in this hypercube, with each point differing from the preceding
one by only one coordinate. In a trajectory, each input parameter
only changes once with pre-defined step $\triangle_{i}$. The function
$y$ (i.e. the simulation model) is evaluated for every point in the
trajectory. The first point of a trajectory is randomly selected.
At each step in a trajectory, the coordinate to be modified is also
randomly selected. Thus various trajectories differ by their starting
point and by the order of modified coordinates. In order to ensure
a strict equal probability for each value, Morris suggested taking
$p$ even and $\Delta_{i}=\frac{p}{2(p_{i}-1)}$. So, if the initial
value $x_{i}$ is lower than 0.5, the final value is $x_{i}+\Delta_{i}$;
if it is greater than 0.5, the final value is $x_{i}-\Delta_{i}$.
Taking this $\Delta_{i}$ value corresponds to a uniform distribution
over the discrete grid values in intervals {[}0;1{]} for all input
variables over the hypercube $H^{k}$.

A trajectory enables a coefficient of variation for each input variable
$i$, called an elementary effect (EE) to be evaluated. It is computed
between the two points of the trajectory where this input variable
$i$ is modified (equation \ref{eq:EEi_Moris})

\begin{equation}
EE_{i}=\frac{\left[y\left(x+e_{i}\triangle_{i}\right)-y\left(x\right)\right]}{\triangle_{i}}\label{eq:EEi_Moris}
\end{equation}
$e_{i}$ is a vector of zeros but with its $i$-th component equal
to $\pm1$. Each trajectory, with its $\left(k+1\right)$ simulations,
provides an estimate of the $k$ elementary effects. A set of $r$
different random trajectories (with index $t$) is defined in the
hypercube of input variables; it provides $r$ estimates $EE_{it}$
of elementary effects related to each input variable $i$, at the
cost of $r\times\left(k+1\right)$ simulations. The average and the
standard deviation of the elementary effects are computed for each
input variable $i$ (equations \ref{eq:Moyen} and \ref{eq:standdev}):

\selectlanguage{english}%
\begin{equation}
\mu_{i}=\frac{1}{r}\sum_{t=1}^{r}\, EE_{it}\label{eq:Moyen}
\end{equation}

\begin{equation}
\sigma_{i}=\sqrt{\frac{1}{\left(r-1\right)}\sum_{t=1}^{r}\left(EE_{it}-\mu_{i}\right)^{2}}\label{eq:standdev}
\end{equation}

\selectlanguage{american}%
Campolongo et al. \citep{CampolongoLARGE} recommend using the average
of absolute elementary effects (equation \ref{eq:Moyen-absolue})
rather than the usual average, since some elementary effects can eliminate
each other in non-monotonic models. 

\selectlanguage{english}%
\begin{equation}
\mu_{i}^{*}=\frac{1}{r}\sum_{t=1}^{r}\,|EE_{it}|\label{eq:Moyen-absolue}
\end{equation}

The criterion \foreignlanguage{american}{$\mu_{i}^{*}$ is a good
indicator to classify input variables by order of importance, despite
the fact that information about the sign of the elementary effects
is lost. Moreover, the standard deviation of the elementary effects
is a relevant indicator of non-linearity in input parameters of the
model or interactions with other parameters involved in the model.
By plotting both statistical indicators, the Morris method identifies
the inputs that can be considered to have an effect : }
\selectlanguage{american}%
\begin{enumerate}
\item Negligible (low average, low standard deviation) 
\item Linear and additive (high average, low standard deviation) 
\item Non-linear or involved in interactions with other input parameters
(high standard deviation). 
\end{enumerate}

\subsection{The second-order interactions sensitivity analysis}

If parameters show a marked non-linear behavior and a significant
influence in the model, a subsequent experiment with only these inputs
is recommended. Campolongo and Braddock \citep{Campolongo19991,Cropp200277}
proposed an extension of the Morris method which enables the calculation
of at least second-order effects with a reasonable computational cost.
This method is based on the calculation of the equivalent of a second-order
derivative of the model and optimizes the computational experiment
by using the solution of the \textquotedblleft{}handcuffed prisoner
problem\textquotedblright{} \citep{Mendelshon1977}. Thus, the number
of evaluations of the model required to obtain one second-order elementary
effect is optimized (about $k^{2}r$). Campolongo and Braddock \citep{Campolongo19991}
defined a second-order elementary effect (equation \ref{eq:Morris_EEij}).

\begin{equation}
EE_{ij}=\left|SEE_{ij}-\frac{EE_{i}}{\triangle_{j}}-\frac{EE_{j}}{\triangle_{i}}\right|\label{eq:Morris_EEij}
\end{equation}

\selectlanguage{english}%
where $SEE_{ij}$ characterizes the influence due to the change of
both factors $i$ and $j$ at the same trajectory (equation \ref{eq:SEEij_morris})

\selectlanguage{american}%
\begin{equation}
SEE_{ij}=\frac{\left[y\left(x+e_{i}\triangle_{i}+e_{j}\triangle_{j}\right)-y\left(x\right)\right]}{\triangle_{i}\triangle_{j}}\label{eq:SEEij_morris}
\end{equation}

\selectlanguage{english}%
A further extension to the analysis of third-order (or higher) interactions
was proposed by Campolongo and Braddock \citep{Campolongo19991}.
Unfortunately the computational cost of this analysis can be prohibitive
if the model is too complex. One practical solution is to perform
a second-order experiment first and then an analysis of third-order
effects only in the group with the highest values of standard deviation
in the second-order analysis. 

Statistics can be applied to second-order elementary effects so that
the results of the mean, absolute mean and standard deviation of the
effects can be interpreted in the same way as in the first-order Morris
method \citep{Campolongo19991}. In addition, as the standard-deviation
of the first-order provides information about second- and higher-order
interactions between the parameters, the second-order extension can
give valuable data about the interactions of third- and upper-order
degrees. The combination of both methods at the same intervals enables
the importance of a parameter in the first, second, and higher orders
to be classified. This combination gives a better interpretation of
complex models and can be applied to the construction of reduced models
based only on the most important parameters.

\selectlanguage{american}%

\subsection{Test case: an apartment building}

\selectlanguage{english}%
In order to explore the potential of the elementary effects method,
a rather complex multi-zone building was chosen as a test case.\foreignlanguage{american}{
It is a seven-storey residential building with 32 dwellings and an
estimated population of between 70 and 80 inhabitants (Figure \ref{fig:The-collective-building}).
It has an approximate floor area of 3500 $m^{2}$, and the average
yearly energy needs for heating are 67.59 $kWh/m^{2}$. The characteristics
of this building are given in Table \ref{tab:thermal-characteristics-of-building-test}.
It is connected to a heating network and no cooling equipment is available.
The glazing ratio is 30\%, composed mainly of double glazed windows.
All exterior walls are insulated with 10 cm of insulating material.
The building is divided into 24 thermal zones (two heated and non-heated
zones per floor).}

\selectlanguage{american}%
The building was modeled using the energy calculation software ESP-r
\citep{Strachan2008601}. The choice of ESP-r was driven by its detailed
sub-models as well as the numerous input variables considered in this
tool, allowing many possible situations of influential factors to
be investigated. The $p$-dimensional grid was defined as $p$=10
for all experiments, so each reduced input variable could take the
discretized values $0,0.111,0.222,...,0.889,1$ in the reduced interval
$[0;1]$ and simulations were made with regularly spaced values between
a minimum and a maximum.

The numerical experiments were classified into two different sets
(A, B) based on the number of parameters involved in the analysis
(see Table \ref{tab:EXP24parameters} for the definition of all input
variables and their range).

In the set of experiments (A), a first-order effects sensitivity analysis
was performed with 24 parameters, representing different choices in
building design. These design parameters were the corrections for
building height (input 1), width (2) and length (3) as well as the
building rotation (24), insulation thickness (23) and glazing ratio
(12-15). Corrections of the weather parameters used in the analysis
were added in a first attempt to quantify the importance of the environment
of the building in the different outputs studied (19-22). 

For some parameters, common values for all heated zones except one
were taken: heat gains due to occupants (8), the set point temperature
(4), the ventilation rate (10) and the difference in the set point
temperature between day and night (6). The objective was to see the
influence of these common parameters compared to the climatic and
design factors that also affect the entire building. In addition,
to determine the relative influence of these parameters in a single
thermal zone, apartment E42 (which is located on the fourth floor)
was chosen arbitrarily to change these parameters only in this specific
zone. However, the specific results from this apartment are not presented
here, as they do not affect the results for the building as a whole
(parameters 5-7-9-11).

For the first numerical experiment (A), rather broad intervals were
chosen for all input variables, representing the diversity of the
characteristics of apartment buildings in an urban context on a large
scale (e.g. at national level), regarding size, insulation, glazing
ratio, and ventilation. In a similar way, broad ranges were taken
for variables concerning the building environment (particularly the
climate) and occupancy. Nevertheless, the values adopted remain representative
of the variations or uncertainties usually considered. It was a deliberate
choice to start with such a ``maximum-variability'' scenario in
order to be representative of apartment buildings at a national level.
One possible application could be to identify the data to be collected
with high priority for a good description of the building stock in
a region or a country.

In the second-order Morris extension experiments (set B), only 12
of 24 parameters from the previous experiment were selected. This
choice was made due to the high computational cost of the analysis.

\section{Results and discussion }

\subsection{General remarks on the presentation of results}

The results of the elementary effects analysis of building thermal
simulations are presented as scatter plots with a point for each input
variable $i$: the x-axis represents the absolute average ($\mu_{i}*$)
and the y-axis represents the standard-deviation of the elementary
effects ($\sigma_{i}$) (see, for example, Figures \ref{fig:figure1_experiments}
to \ref{fig:FIGURE5}). 

The absolute average ($\mu_{i}^{*}$) was introduced above as a measure
of importance for the input factor $i$. This information can be complemented
by the ratio ($\sigma_{i}/\mu_{i}^{*}$) as an indicator of linearity
(or non-linearity), as justified below.

First, looking at one input variable $i$ : if all estimates of elementary
effects $EE_{it}$ have the same sign, we can say that this input
factor $i$ has a monotonic effect on the response $y$, increasing
or decreasing, depending on the sign of the elementary effects. In
this case, $\mu_{i}^{*}$ is equal to the absolute value of $\mu_{i}$.
The reverse is also true : if $\mu_{i}^{*}=abs\left(\mu_{i}\right)$,
the effects of input variable i are monotonic.

Using well-known statistical properties, if elementary effects are
assumed to be normally distributed, 95\% of $EE$-estimates are within
the range $\left(\mu_{i}\pm1.96\,\sigma_{i}\right)$. As a consequence,
if \foreignlanguage{english}{$\sigma_{i}/\mu_{i}$} is smaller than
0.10, most elementary effects (95\%) are in a range \foreignlanguage{english}{$\pm20\%$}
around \foreignlanguage{english}{$\mu_{i}$}: the elementary effects
are almost constant and the input variable $i$ has an almost linear
effect on the model. A true linear response correspond to $\sigma_{i}/\mu_{i}=0$.

If the ratio \foreignlanguage{english}{$\sigma_{i}/\mu_{i}$} is smaller
than 0.5, most elementary effects (95\% with the normal assumption)
have the same sign and the model response can be considered as monotonic
with respect to the input factor $i$. Thus, in this case, it can
be considered that $\mu_{i}\thickapprox abs\left(\mu_{i}\right)$
and $\sigma_{i}/\mu_{i}^{*}\thickapprox\sigma_{i}/abs\left(\mu_{i}\right)$.
This fact justifies using the ratio \foreignlanguage{english}{$\sigma_{i}/\mu_{i}^{*}$}
as an indicator for almost linear (if <0.1) or monotonic influences
(if < 0.5).

As the distribution of elementary effects may be far from normal,
another justification is presented in Figure \ref{fig:Scatterplot}
with a scatter plot of \foreignlanguage{english}{$\left(\sigma_{i}/\mu_{i}^{*}\right)$}
vs. \foreignlanguage{english}{$\sigma_{i}/abs\left(\mu_{i}\right)$}
for all first-order analyzes performed within the present work (sets
A and B1, B2). This diagram shows that for \foreignlanguage{english}{$\sigma_{i}/abs\left(\mu_{i}\right)$}
smaller than one, most of the points are on the bisector \foreignlanguage{english}{$\left(\sigma_{i}/\mu_{i}^{*}\right)=\sigma_{i}/abs\left(\mu_{i}\right)$},
indicating a monotonic (or almost monotonic) behavior in a wider interval
than expected (not only < 0.5). For highly scattered elementary effects
\foreignlanguage{english}{$\left(\sigma_{i}/abs\left(\mu_{i}\right)>1\right)$},
a non-monotonic behavior is clearly established and, in this case,
\foreignlanguage{english}{$\left(\sigma_{i}/\mu_{i}^{*}\right)$}
stays in the interval between 1 and 2. The absolute average \foreignlanguage{english}{$\mu_{i}^{*}$}
is very different from the average \foreignlanguage{english}{$abs\left(\mu_{i}\right)$}
and is more influenced by the standard deviation $\sigma_{i}$.

So, by plotting three straight lines of slopes \foreignlanguage{english}{$\sigma/\mu^{*}=$}0.1,
0.5 and 1, respectively, we can graphically identify in the elementary
effects scatter plot, those factors which are almost linear (below
the line \foreignlanguage{english}{$\sigma/\mu^{*}=$} 0.1), monotonic
\foreignlanguage{english}{(0.5 > $\sigma/\mu^{*}$> 0.1)} or almost
monotonic $\left(1>\sigma/\mu^{*}>0.5\right)$, and those factors
with marked non-monotonic non-linearities or interactions with other
factors $\left(\sigma/\mu^{*}>1\right)$.

Defining these four zones also provides a means of checking the results
of the sensitivity analysis if the results contradict what is understood
from the physical point of view. 

For some of the figures (for example Figs. \ref{fig:4c}, \ref{fig:4d},
\ref{fig:5c} and \ref{fig:5d}), a slightly different presentation
of data has been used with the ratio \foreignlanguage{english}{$\sigma_{i}/\mu_{i}^{*}$}
on the vertical axis (the different domains defined above as linear,
monotonic and highly non-linear being distinguished by horizontal
lines). This presentation provides exactly the same information but
is sometimes easier to read as the different points are better separated.

\subsection{Computational experiment A: first-order analysis results}

For the first set of numerical experiments (set A), several characteristics
for output results were examined using the first-order elementary
effects analysis: yearly heating loads (Fig. \ref{fig:1a}), outputs
derived from the latter (heating load per m$^{3}$; logarithmic transformation)
(Fig. \ref{fig:1b} - \ref{fig:1d}), heating power (with the example
of the power exceeded during 1000 hours/year) (Fig. \ref{fig:2c})
and the summer comfort factor, using the average internal temperature
in July-August (Fig. \ref{fig:2a}).

Before presenting the results, it is important to recall the definition
of elementary effects: one value of the elementary effect for the
input variable $i$ corresponds to the output variation when the input
$i$ moves from the minimum (0) to the maximum (1). For example, in
Figure \ref{fig:1a}, the average elementary effect for the building
height (input 1) has a value of 470 000 kWh/year. This is the average
influence of a modification of building height from 9 m to 27 m (nominal
height is 18 m, corrected by $\pm$50\%). For the building considered
and the wide intervals taken for input parameters, the yearly energy
needs are within the range of 57 587 kWh/year to 788 894 kWh/year.

The first analysis based on the yearly heating load (Fig. \ref{fig:1a})
highlights the dimensional parameters (numbered 1, 2 and 3), indicating
a typical size effect: building size is the major influence on heating
demand. Next, a second group of input parameters appears with a significant
influence: set point temperature (4), ventilation rate (10) and insulation
thickness (23). The remaining parameters can be classified as less
important, although not negligible for a number of them (6, 8, 12,
13, 19, 20, 21, 24) which will be discussed below in this section.
Only one parameter has an almost linear effect (occupant free heat
gains, 8) with $\sigma/\mu*$ close to 0.1. All other parameters show
a non-linear influence and/or interactions with other parameters ($\sigma/\mu*$
> 0.5 for many of them). Nevertheless, they generally stay within
the monotonic zone ($\sigma/\mu*$< 1) with the notable exception
of insulation thickness (23).

An attempt to eliminate the size effect is made through the analysis
of annual heating load per cubic meter, i.e. load divided by the three-dimensional
parameters (Fig. \ref{fig:1c}). The second group of parameters identified
in the former case (4, 10 and 23) now becomes predominant, with a
behavior closer to linear for the set point temperature (4) and ventilation
rate (10) ($\sigma/\mu*$ equals 0.3 and 0.2, respectively).

A third group of parameters is emphasized by this presentation, including
factors depending on occupants (free heat gains, 8, and heating set
point night reduction, 6), solar gains (diffuse, 19, and direct, 21,
radiation potential; building rotation, 24; glazing ratios on main
facades, 12 and 13) and climatic sensitivity related to the building
environment (external temperature, 20). Compared with Fig \ref{fig:1a},
all these parameters coincide with those already identified as secondary
but non-negligible.

The importance of dimensional parameters (1, 2 and 3) is clearly reduced,
but they remain among the non-negligible variables and contribute
to the high variability of the model ($\sigma/\mu*$ close to 1).

A logarithmic transformation is applied to the output to identify
interactions caused by possible multiplicative effects (Figure \ref{fig:1b}
for annual heating load and \ref{fig:1d} for heating load per cubic
meter). Logarithmic transformation is often used for the statistical
analysis of positive outputs with variations of several orders of
magnitude or when multiplicative phenomena take place. If the response
function can be expressed as a multiplicative function of various
inputs $x_{1}$, $x_{2}$, $x_{3}$... as $y(x)$$=f_{1}(x_{1})\times f_{2}(x_{2})\times f_{3}(x3)...$,
the elementary effects of $y$ vs. input variables $x_{1}$, $x_{2}$,
$x_{3}$... are highly influenced by interactions between $x_{1}$,
$x_{2}$, $x_{3}$... The standard deviations of their respective
elementary effects can be expected to have high values. Taking the
logarithm of $y(x)$ separates the terms $ln(y(x))=ln(f_{1})+ln(f_{2})+ln(f_{3})...$
and eliminates interactions: for the logarithm output, the standard
deviation of EE depends only on the curvature of function $ln(f_{1});ln(f_{2});ln(f_{3})...$
Another advantage of the natural logarithmic transformation is the
interpretation of the elementary effects as rates of change of the
output when the value remains low (typically <0.5). For instance,
regarding the input variable \emph{'night and day difference temperature'}
(6), the average elementary effect (expressed in logarithm) is 0.35
(here a decrease, i.e. -0.35, Figure \ref{fig:1d}). This means that
a full-scale change in this input (from 0 K to 8 K) leads to a 35\%
variation in space heating, on average.

The logarithmic transformation provides consistent results with the
former ones: the same input parameters are identified as important
and non-important with about the same ranking and similar clusters.
The use of a logarithmic transformation emphasizes some factors with
partial multiplicative effects, mainly temperature-related parameters:
set point (4), night reduction (6) and external temperature (20).
Their $\sigma/\mu*$ ratios are both reduced significantly when compared
with Fig. \ref{fig:1a} and \ref{fig:1b}, and are close to 0.1.

Other parameters remain non-linear with the logarithmic transformation,
such as parameters related to the glazing ratio (12, 13) and rotation
(24) or insulation thickness (23). It can be noted that the two presentations
(Fig. \ref{fig:1b} and \ref{fig:1d}) in the logarithmic transformation
lead to exactly the same coordinates for all input parameters, except
the dimensional ones (1, 2 and 3).

As a partial conclusion, all the presentations show consistent results
in terms of predominant parameters. The high variability of almost
all the parameters encourages the use of a methodology to analyze
the effects of interactions and non-linearities. It is notable that
the most important parameters in the analysis (ventilation and set
point temperature) are in accordance with a similar analysis carried
out by Brohus et al. \citep{Brohus2009} for a single-family detached
house model with 71 input parameters. The glazing ratio appears to
be less important for the model output of energy demand for heating.
Similar conclusions were also reported by Gasparella et al. \citep{Gasparella20111030}
who did an analysis and modeling of various types of glazing in a
high insulated building with a high percentage of double glazing.

The Morris method analysis was applied to the average internal temperature
in July and August (Figure \ref{fig:2a}) and to the power in the
monotonic curve at hour 1000 (Figure \ref{fig:2c}). The objective
was to show the applicability of this method to different kinds of
output. By comparing Figures \ref{fig:1b} and \ref{fig:2c}, we can
conclude that there is a strong relationship between the output of
heating needs in ln(kWh/year) and the monotonic curve of power. Only
the set point temperature difference between night and day (6) becomes
less important in terms of average elementary effects but with a higher
standard deviation. On the contrary, the comfort criterion (Figure
\ref{fig:2a}) shows a very different behavior. The most important
parameter is related to the external temperature with a highly linear
effect. The next important parameters (with a high $\mu_{i}*$) are
related to the environment or the geometry of the building. According
to Figure \ref{fig:2a}, the glazing ratio of side B (parameter 13)
seems to have more importance in the variability of the average temperature
than the other sides (parameters 12, 14 and 15). This implies that
parameter 13 could have a major impact on the comfort level of the
building. The same conclusions can be made by analyzing the geometrical
parameters (1, 2 and 3). Thus, the Morris method could also serve
as a diagnostic tool to help designers to review the impact of the
building on comfort levels and to analyze in more depth the sensitivity
of the building in terms of solar gains.

\subsection{Computational experiment B: first- and second-order results}

\subsubsection{Data used in experiment B}

In order to make the analysis done with the first-order approach in
experiment A more precise, experiments B1 to B4 were performed, focusing
on the study of the second-order interactions. Due to the high computational
cost of second order analysis (5760 runs with $r=10$), a subset of
only 12 from 24 initial input parameters was selected (see Table \ref{tab:EXP24parameters}).
Parameters with negligible influence were omitted and important parameters
(from experiment A) were kept with exceptions for some parameters
playing similar roles: among size-related parameters (1, 2, 3), the
height was omitted and among the glazing ratio of the facades (11,
12, 13, 14), only the one for facade A (11) was kept. Furthermore,
the analysis was done for two interval sizes for the parameters: a
large one (the same as the one used previously for the first-order
experiment) and a small one (Table 2). These small intervals may be
considered representative of the level of knowledge or uncertainties
of the main parameters in the situation of a detailed energy audit
(with measurements being made inside the building). Small intervals
can be considered for parameters known approximately while large intervals
can be used for totally unknown parameters. 

Two different outputs are presented: the annual heating needs and
the natural logarithm of the annual heating needs per cubic meter.
Before running the second-order experiment, a first-order analysis
is made, limited to the 12 input variables selected. So experiment
B actually incorporates 4 numerical experiments: B1 (1st order, large
interval, same interval as experiment A), B2 (2nd order, large interval),
B3 (1st order, small interval) and B4 (2nd order, small interval).

\subsubsection{First order - large interval (B1)}

A comparison between the two first-order experiments B1 (12 parameters:
Fig. \ref{fig:4a} and \ref{fig:4b}) and A (24 parameters: Fig. \ref{fig:1a}
and \ref{fig:1d}) demonstrates that the decrease or increase in the
number of parameters involved in the sensitivity analysis does not
significantly affect the ranking between the parameters. When comparing
the two sets of results, one should keep in mind that differences
may come from two sources:

- the Morris Method is basically a Monte Carlo approach with random
generation, so changes may come from sampling differences,

- reducing the number of variables (with unchanged intervals) results
in fewer interactions between input variables, so a lower standard
deviation for elementary effects can be expected. This trend is actually
observed for most of the variables for the two output responses considered,
but with a limited decrease in the standard deviation.

\subsubsection{Second order (B2)}

For the second-order analysis (experiment B2, Fig. \ref{fig:4c} and
\ref{fig:4d}), it can be seen by comparing the x-axis of Figures
\ref{fig:4a} and \ref{fig:4c} that the second-order influences are
not negligible. Figure \ref{fig:4c} (with logarithmic response) shows
that the average of the second-order elementary effect of the highest
influence interaction (3;8) is almost 0.2, exceeding some significant
first-order influences of parameters (e.g. 2 and 24) (Figure \ref{fig:4a}).
This influence becomes even more evident in the analysis of annual
heating needs in kWh (Figures \ref{fig:4b} and \ref{fig:4d}). The
average second-order elementary effect of the highest influence (3,2)
exceeds first-order averages of elementary effects in all parameters
with the exception of parameters 2, 3, and 4.

Size parameters must be emphasized as being the main source of interactions;
width (2) and length (3) are included in the five highest second-order
elementary effects in the two presentations. For the annual heating
load (Fig. \ref{fig:4d}), their own interaction (2-3) is the main
one followed by their interaction with the set point temperature (4)
and the ventilation rate (10). These results are fully consistent
with the first-order analysis (Fig. \ref{fig:4b}) which pinpoints
parameters 2, 3, 4, 10 and 23 (insulation) as the major influences
and highly scattered elementary effects. In the logarithmic transformation,
insulation thickness is clearly the factor with the highest sigma
values (highly scattered elementary effects) so it is not surprising
to see interactions with size parameters among the main second-order
interactions. High interactions with occupant heat gain (8) in effects
(3; 8) and (2; 8) are more surprising. A possible interpretation could
be that free heat gains have an almost additive influence on the space
heating load as indicated by the fact that parameter 8 is the only
one in the linear zone in the first-order analysis (Figs. \ref{fig:1a}
and \ref{fig:4b}). Dividing by the building volume introduces high
interactions with size parameters while the logarithmic transformation
introduces non-linearity for such additive phenomena: with the different
transformations (Fig. \ref{fig:1b}, \ref{fig:1c} and \ref{fig:1d})
input parameter 8 is no longer in the linear zone. No other factors
are involved in the significant second-order interactions with occupants.
Taking into account the size effect of occupants would suggest using
$occupants/m^{3}$ (or $occupants/m^{2}of-floor-area$)  as an input
parameter instead of just occupants in further work.

Regarding the possible linearity of second order effects, no pair
of parameters in Figures \ref{fig:4c} and \ref{fig:4d} are in the
linear zone $\left(\sigma/\mu*<0.1\right)$, which means that all
pairs of parameters have non-linear effects or are combined with other
parameters in the higher order. This can be compared to the first-order
analysis for which only parameter 4 (set point temperature) falls
in the almost linear zones (Fig. \ref{fig:4a}). Continuing the analysis
of parameter 4, Figure \ref{fig:4c} emphasizes the strong interactions
of this parameter with the size of the building (parameters 2 and
3). These interactions remain important even if the size effect is
reduced by considering the heating needs per cubic meter as the output.
Parameter 23 (insulation thickness of external walls) appears in all
experiments (Figure \ref{fig:FIGURE4}) with high variability in the
first order as well as in the second order. The non-linear influence
of the insulation thickness can explain this high variability. 

In contrast, the first-order elementary effect of set point temperature
(4) (non-monotonic for the annual heating needs) moves to the linear
zone after logarithmic transformation per cubic meter, with the highest
influence. The standard deviation remains non-negligible if compared
with other factors and a second-order interaction with size parameters
appears even with the logarithm, but other interactions are limited.

\subsubsection{First order - small interval (B3)}

The same kind of analysis in Figure \ref{fig:FIGURE4} can be found
in Figure \ref{fig:FIGURE5} where a small interval is considered.
The main consequence of reducing the interval size of parameters (by
a factor 10 for most of input parameters) is that almost all parameters
are now in the linear zone for the first and second order. The set
point temperature is again the most important parameter in this experiment
with the highest average and standard deviation of the first-order
elementary effects (all outputs in Fig. \ref{fig:FIGURE5}). The importance
of set point temperature is highlighted by the fact that its variation
interval is kept rather wide in the experiment (reduction by a factor
4 only if compared to the large interval). In the reverse, insulation
thickness has now almost no influence on the output, due to a choice
of rather well insulated wall (between 9 and 10 cm) for which the
precise thickness is of little importance.

\subsubsection{Second order - small interval (B4)}

As first-order analysis results show almost a linear behavior for
most of input parameters, one could consider that second-order analysis
could be of little interest. Actually, results can be considered as
``second-order\textquotedblright{} magnitude : Second-order average
elementary effects are smaller than 0.3\% in the logarithmic output,
e.g.. Moreover, they are close to the standard deviation of the first
order analysis in the two presentations -kWh or Ln($kWh/m^{3}$)-,
showing that scattering of first order EE are mainly caused by interactions
between input parameters and not by quadratic or other non-linear
influence. For the yearly energy output (kWh; Fig. \ref{fig:5d}),
the four factors dominating the first order analysis (2, 3, 4 and
10) are encountered again: the six main second-order EEs correspond
to their six pairs of double interactions, with interactions between
size parameters (2, 3) and set-point (4) being the largest values.
In the logarithmic output (Fig. \ref{fig:5c}), the interactions between
size parameters (2, 3) and occupant heat gains (8) analyzed in section
4.3.3 for the large interval are encountered again. Size parameters
are not involved in any significant interactions. But other interactions
involving the set point temperature (4), the most influential parameter,
are also identified with night-reduction set point (6) and occupant
heat gains (8). For the first time in our analysis, input parameters
related to a specific thermal zone of the building are shown as non
negligible, with interaction between set point temperature in the
whole building (4) and in this specific zone (5). Significant heat
transfer between zones could explain this phenomenon which could become
significant in situations where building characteristics are precisely
known and important uncertainties could be related to diverse inhabitant
behaviors in the various zones of the building. Standard deviations
of second order elementary effects (in the logarithmic output) are
always smaller than 0.02\%, so third- or upper-order analysis will
bring no useful information.

\section{Conclusions}

In this work, first- and second-order sensitivity analyzes have been
combined. The Morris method and its extension have been applied to
a building thermal simulation case study, using ESP-r to calculate
the output. Among the results of this approach, the first order-analysis
has demonstrated that, even if a precise ranking between the input
parameters is not relevant, they can be split into different families
in order to discuss their importance. This first order analysis helps
to identify possible non-linearity or interactions of higher orders.
The usefulness of considering various forms for the output function
(kWh/year, kWh/year.m$^{3}$, ln(kWh/year.m$^{3}$)) has been explored,
in particular to reduce the number of variables affected by these
non-linearities or interactions. For instance, specific values per
m$^{3}$ (or per m$^{2}$) reduce the correlation to the size-related
parameters. Moreover, considering the logarithm of the output helps
to identify the origin of some of the non-linearity. It is worth noting
that investigating various model outputs does not require additional
simulation runs: the same set of simulation trajectories is used and
only complementary post-processing is needed. This remark can be extended
to any transformed output calculated from model output and input variables.
For those parameters still remaining in the area of high standard
deviation, it has been demonstrated how the implementation of the
second-order sensitivity analysis can be helpful to sort variables
and to specify their interaction in pairs. It has also been shown
how the usefulness of the upper-order analysis is amplified when combined
with the different forms for the output. A new way of presenting results
from the Morris method has been proposed to classify parameters (or
couples of parameters in the case of second-order analysis) according
to their sensitivity: linear- , monotonic- , almost monotonic or highly
non-linear/interaction-of-higher-order). The value of sensitivity
analysis using the elementary effects method has been clearly established.
In any case, for a given building being simulated with a specific
modeling tool, no general sensitivity can be derived: the results
depend on the input parameters, with fixed values or varying values
(and for the latter, their variation range). Thus, sensitivity analysis
must be performed for each particular situation, in relation to modeling
goals, with a careful choice of variation intervals. Its conclusions
are valid only for this particular situation.

The choice of variation interval for each input parameter must be
related to either a constrained range for decision variables or an
uncertainty domain for exogenous variables.

\section*{Acknowledgments.}

This work was funded by the region of Pays de la Loire (France) in
the framework of the research program MEIGEVILLE, coordinated by the
IRSTV (\emph{Institut de Recherche en Science et Technique de la Ville,
FR-CNRS 2488}).

\bibliographystyle{elsarticle-num}
\bibliography{ch3,paper1}

\section*{List of figures.}

Fig. 1. The apartment building case study.

Fig. 2. Scatter plot of $\left(\sigma_{i}/\mu_{i}^{*}\right)$ vs.
$\left(\sigma_{i}/abs\left(\mu_{i}\right)\right)$ for all first-order
analyzes performed (sets A1,B1,B2).

Fig. 3. Estimated absolute average ($\mu*$) and standard deviation
($\sigma$) of the first-order elementary effects for different sets
of outputs related to energy consumption (see Table \ref{tab:EXP24parameters},
experiment A1). The numbers on the plot are the parameter indices
and the lines represent the slope $\sigma/\mu*$ at the values 0.1,
0.5 and 1. 

Fig. 4. Estimated absolute average ($\mu*$) and standard deviation
($\sigma$) of the first-order elementary effects in experiments for
different sets of outputs related to power and thermal comfort (see
Table \ref{tab:EXP24parameters}, experiment A1). The number in the
plot are the parameter indices and the lines represent the slope $\sigma/\mu*$
at the values 0.1, 0.5 and 1. 

Fig. 5. Estimated first- and second-order elementary effects for two
outputs related to annual heating needs in kWh and in ln(kWh/year/m$^{3}$)
with a large interval between the parameters and $r=10$. Both methodologies
have the same number of parameters (Table \ref{tab:EXP24parameters},
experiments B1 and B2) The lines represent the slope $\sigma/\mu*$
at the values 0.1, 0.5 and 1.

Fig. 6. Estimated first- and second-order elementary effects for two
outputs related to annual heating needs in kWh and in ln(kWh/year/m$^{3}$)
with a small interval between the parameters and $r=10$ (table \ref{tab:EXP24parameters}).
The lines represent the slope $\sigma/\mu*$ at the values 0.1, 0.5
and 1.

\section*{Tables}

\begin{center}
\begin{table}[H]
\selectlanguage{british}%
\begin{centering}

\resizebox{.50\textwidth}{!}{
\begin{tabular}{>{\raggedright}m{0.49\textwidth}r}
\hline 
\selectlanguage{american}%
Year built (Base case)\selectlanguage{british}%
 & \selectlanguage{american}%
1990\selectlanguage{british}%
\tabularnewline
\hline 
\selectlanguage{american}%
Floor area \foreignlanguage{english}{$\textrm{\ensuremath{\left(m^{2}\right)}}$}\selectlanguage{british}%
 & \selectlanguage{american}%
3500\selectlanguage{british}%
\tabularnewline
\selectlanguage{american}%
Width (m)\selectlanguage{british}%
 & \selectlanguage{american}%
14.00\selectlanguage{british}%
\tabularnewline
\selectlanguage{american}%
Length (m)\selectlanguage{british}%
 & \selectlanguage{american}%
33.40\selectlanguage{british}%
\tabularnewline
\selectlanguage{american}%
Height (m)\selectlanguage{british}%
 & \selectlanguage{american}%
18.00\selectlanguage{british}%
\tabularnewline
\selectlanguage{american}%
U-value external walls $\textrm{\ensuremath{\left(\mathsf{W/m^{2}K}\right)}}$\selectlanguage{british}%
 & \selectlanguage{american}%
0.472\selectlanguage{british}%
\tabularnewline
\selectlanguage{american}%
U-value internal walls $\textrm{\ensuremath{\left(\mathsf{W/m^{2}K}\right)}}$\selectlanguage{british}%
 & \selectlanguage{american}%
4.400\selectlanguage{british}%
\tabularnewline
\selectlanguage{american}%
U-value double-glazing $\textrm{\ensuremath{\left(\mathsf{W/m^{2}K}\right)}}$\selectlanguage{british}%
 & \selectlanguage{american}%
2.811\selectlanguage{british}%
\tabularnewline
\selectlanguage{american}%
U-value basement floor $\textrm{\ensuremath{\left(\mathsf{W/m^{2}K}\right)}}$\selectlanguage{british}%
 & \selectlanguage{american}%
0.887\selectlanguage{british}%
\tabularnewline
\selectlanguage{american}%
Glazing ratio (\%)\selectlanguage{british}%
 & \selectlanguage{american}%
30\selectlanguage{british}%
\tabularnewline
\selectlanguage{american}%
Number of occupants\selectlanguage{british}%
 & \selectlanguage{american}%
70-80\selectlanguage{british}%
\tabularnewline
\hline 
\end{tabular}}
\par\end{centering}

\selectlanguage{american}%
\caption{Main characteristics of the apartment building test.\label{tab:thermal-characteristics-of-building-test}}
\end{table}

\par\end{center}

\selectlanguage{english}%
\begin{center}
\begin{sidewaystable}
\selectlanguage{british}%
\begin{centering}

\resizebox{1\textwidth}{!}{
\begin{tabular*}{1\paperwidth}{@{\extracolsep{\fill}}llcccc}
\toprule 
\multirow{2}{*}{\selectlanguage{american}%
N\selectlanguage{british}%
} & \multirow{2}{*}{\selectlanguage{english}%
Parameter\selectlanguage{british}%
} & \multicolumn{4}{c}{\selectlanguage{english}%
Intervals\selectlanguage{british}%
}\tabularnewline
\cmidrule{3-6} 
 &  & \selectlanguage{american}%
1 (large)\selectlanguage{british}%
 & \selectlanguage{american}%
Experiment\selectlanguage{british}%
 & \selectlanguage{american}%
2 (small)\selectlanguage{british}%
 & \selectlanguage{american}%
Experiment\selectlanguage{british}%
\tabularnewline
\midrule
\selectlanguage{english}%
1\selectlanguage{british}%
 & \selectlanguage{english}%
Building size: correction for height {[}\%{]}\selectlanguage{british}%
 & \selectlanguage{english}%
50-150\selectlanguage{british}%
 & \selectlanguage{american}%
A\selectlanguage{british}%
 & \selectlanguage{american}%
\selectlanguage{british}%
 & \selectlanguage{american}%
\selectlanguage{british}%
\tabularnewline
\selectlanguage{english}%
2\selectlanguage{british}%
 & \selectlanguage{english}%
Building size: correction for width {[}\%{]}\selectlanguage{british}%
 & \selectlanguage{english}%
50-150\selectlanguage{british}%
 & \selectlanguage{american}%
A,B1,B2\selectlanguage{british}%
 & \selectlanguage{english}%
90-100\selectlanguage{british}%
 & \selectlanguage{american}%
B3,B4\selectlanguage{british}%
\tabularnewline
\selectlanguage{english}%
3\selectlanguage{british}%
 & \selectlanguage{english}%
Building size: correction for length {[}\%{]}\selectlanguage{british}%
 & \selectlanguage{english}%
50-150\selectlanguage{british}%
 & \selectlanguage{american}%
A,B1,B2\selectlanguage{british}%
 & \selectlanguage{english}%
90-100\selectlanguage{british}%
 & \selectlanguage{american}%
B3,B4\selectlanguage{british}%
\tabularnewline
\selectlanguage{english}%
4\selectlanguage{british}%
 & \selectlanguage{english}%
Set point temperature of all apartments except Apt.E42 {[}\textdegree{}C{]}\selectlanguage{british}%
 & \selectlanguage{english}%
17-24\selectlanguage{british}%
 & \selectlanguage{american}%
A,B1,B2\selectlanguage{british}%
 & \selectlanguage{english}%
20-22\selectlanguage{british}%
 & \selectlanguage{american}%
B3,B4\selectlanguage{british}%
\tabularnewline
\selectlanguage{english}%
5\selectlanguage{british}%
 & \selectlanguage{english}%
Set point temperature of Apt. E42 {[}\textdegree{}C{]}\selectlanguage{british}%
 & \selectlanguage{english}%
17-24\selectlanguage{british}%
 & \selectlanguage{american}%
A,B1,B2\selectlanguage{british}%
 & \selectlanguage{english}%
20-22\selectlanguage{british}%
 & \selectlanguage{american}%
B3,B4\selectlanguage{british}%
\tabularnewline
\selectlanguage{english}%
6\selectlanguage{british}%
 & \selectlanguage{english}%
Night-day set point temp. diff. affecting all apartments except Apt.
E42 {[}\textdegree{}C{]}\selectlanguage{british}%
 & \selectlanguage{english}%
0-8\selectlanguage{british}%
 & \selectlanguage{american}%
A,B1,B2\selectlanguage{british}%
 & \selectlanguage{english}%
0-1\selectlanguage{british}%
 & \selectlanguage{american}%
B3,B4\selectlanguage{british}%
\tabularnewline
\selectlanguage{english}%
7\selectlanguage{british}%
 & \selectlanguage{english}%
Night-day set point temperature of Apt. E42 {[}\textdegree{}C{]}\selectlanguage{british}%
 & \selectlanguage{english}%
0-8\selectlanguage{british}%
 & \selectlanguage{american}%
A\selectlanguage{british}%
 & \selectlanguage{american}%
\selectlanguage{british}%
 & \selectlanguage{american}%
\selectlanguage{british}%
\tabularnewline
\selectlanguage{english}%
8\selectlanguage{british}%
 & \selectlanguage{english}%
Occupants affecting all apartments except Apt. E42 {[}occ./apt.{]}\selectlanguage{british}%
 & \selectlanguage{english}%
1-8\selectlanguage{british}%
 & \selectlanguage{american}%
A,B1,B2\selectlanguage{british}%
 & \selectlanguage{english}%
3-4\selectlanguage{british}%
 & \selectlanguage{american}%
B3,B4\selectlanguage{british}%
\tabularnewline
\selectlanguage{english}%
9\selectlanguage{british}%
 & \selectlanguage{english}%
Occupants of Apartment. E42 {[}occ./apt.{]}\selectlanguage{british}%
 & \selectlanguage{english}%
1-8\selectlanguage{british}%
 & \selectlanguage{american}%
A\selectlanguage{british}%
 & \selectlanguage{american}%
\selectlanguage{british}%
 & \selectlanguage{american}%
\selectlanguage{british}%
\tabularnewline
\selectlanguage{english}%
10\selectlanguage{british}%
 & \selectlanguage{english}%
Ventilation rate affecting all apartments except Apt. E42 {[}\%{]}\selectlanguage{british}%
 & \selectlanguage{english}%
40-100\selectlanguage{british}%
 & \selectlanguage{american}%
A,B1,B2\selectlanguage{british}%
 & \selectlanguage{english}%
80-90\selectlanguage{british}%
 & \selectlanguage{american}%
B3,B4\selectlanguage{british}%
\tabularnewline
\selectlanguage{english}%
11\selectlanguage{british}%
 & \selectlanguage{english}%
Ventilation rate of Apartment E42 {[}\%{]}\selectlanguage{british}%
 & \selectlanguage{english}%
40-100\selectlanguage{british}%
 & \selectlanguage{american}%
A\selectlanguage{british}%
 & \selectlanguage{american}%
\selectlanguage{british}%
 & \selectlanguage{american}%
\selectlanguage{british}%
\tabularnewline
\selectlanguage{english}%
12\selectlanguage{british}%
 & \selectlanguage{english}%
Glazing ratio A {[}\%{]}\selectlanguage{british}%
 & \selectlanguage{english}%
5-50\selectlanguage{british}%
 & \selectlanguage{american}%
A,B1,B2\selectlanguage{british}%
 & \selectlanguage{english}%
45-50\selectlanguage{british}%
 & \selectlanguage{american}%
B3,B4\selectlanguage{british}%
\tabularnewline
\selectlanguage{american}%
13\selectlanguage{british}%
 & \selectlanguage{american}%
Glazing ratio B {[}\%{]}\selectlanguage{british}%
 & \selectlanguage{english}%
5-50\selectlanguage{british}%
 & \selectlanguage{american}%
A\selectlanguage{british}%
 & \selectlanguage{american}%
\selectlanguage{british}%
 & \selectlanguage{american}%
\selectlanguage{british}%
\tabularnewline
\selectlanguage{american}%
14\selectlanguage{british}%
 & \selectlanguage{american}%
Glazing ratio C {[}\%{]}\selectlanguage{british}%
 & \selectlanguage{english}%
5-50\selectlanguage{british}%
 & \selectlanguage{american}%
A\selectlanguage{british}%
 & \selectlanguage{american}%
\selectlanguage{british}%
 & \selectlanguage{american}%
\selectlanguage{british}%
\tabularnewline
\selectlanguage{american}%
15\selectlanguage{british}%
 & \selectlanguage{american}%
Glazing ratio D {[}\%{]}\selectlanguage{british}%
 & \selectlanguage{english}%
5-50\selectlanguage{british}%
 & \selectlanguage{american}%
A\selectlanguage{british}%
 & \selectlanguage{american}%
\selectlanguage{british}%
 & \selectlanguage{american}%
\selectlanguage{british}%
\tabularnewline
\selectlanguage{english}%
16\selectlanguage{british}%
 & \selectlanguage{english}%
Ground reflectivity {[}\%{]}\selectlanguage{british}%
 & \selectlanguage{english}%
20- 30\selectlanguage{british}%
 & \selectlanguage{american}%
A\selectlanguage{british}%
 & \selectlanguage{american}%
\selectlanguage{british}%
 & \selectlanguage{american}%
\selectlanguage{british}%
\tabularnewline
\selectlanguage{english}%
17\selectlanguage{british}%
 & \selectlanguage{english}%
Ground reflectivity in presence of snow (January-December) {[}\%{]}\selectlanguage{british}%
 & \selectlanguage{english}%
30-50\selectlanguage{british}%
 & \selectlanguage{american}%
A\selectlanguage{british}%
 & \selectlanguage{american}%
\selectlanguage{british}%
 & \selectlanguage{american}%
\selectlanguage{british}%
\tabularnewline
\selectlanguage{english}%
18\selectlanguage{british}%
 & \selectlanguage{english}%
View factor of ground {[}\%{]}\selectlanguage{british}%
 & \selectlanguage{english}%
30-40\selectlanguage{british}%
 & \selectlanguage{american}%
A\selectlanguage{british}%
 & \selectlanguage{american}%
\selectlanguage{british}%
 & \selectlanguage{american}%
\selectlanguage{british}%
\tabularnewline
\selectlanguage{english}%
19\selectlanguage{british}%
 & \selectlanguage{english}%
Climatic sensitivity:correction for horizontal diffuse solar rad.
{[}-{]}\selectlanguage{british}%
 & \selectlanguage{english}%
0.2-1\selectlanguage{british}%
 & \selectlanguage{american}%
A,B1,B2\selectlanguage{british}%
 & \selectlanguage{english}%
0.9-1\selectlanguage{british}%
 & \selectlanguage{american}%
B3,B4\selectlanguage{british}%
\tabularnewline
\selectlanguage{english}%
20\selectlanguage{british}%
 & \selectlanguage{english}%
Climatic sensitivity:correction for external dry bulb temp. {[}-{]}\selectlanguage{british}%
 & \selectlanguage{english}%
0.2-1\selectlanguage{british}%
 & \selectlanguage{american}%
A\selectlanguage{british}%
 & \selectlanguage{american}%
\selectlanguage{british}%
 & \selectlanguage{american}%
\selectlanguage{british}%
\tabularnewline
\selectlanguage{english}%
21\selectlanguage{british}%
 & \selectlanguage{english}%
Climatic sensitivity:correction for direct normal solar intensity
{[}-{]}\selectlanguage{british}%
 & \selectlanguage{english}%
0.2-1\selectlanguage{british}%
 & \selectlanguage{american}%
A,B1,B2\selectlanguage{british}%
 & \selectlanguage{english}%
0.2-1\selectlanguage{british}%
 & \selectlanguage{american}%
B3,B4\selectlanguage{british}%
\tabularnewline
\selectlanguage{english}%
22\selectlanguage{british}%
 & \selectlanguage{english}%
Climatic sensitivity:correction for wind speed {[}-{]}\selectlanguage{british}%
 & \selectlanguage{english}%
0.5-1\selectlanguage{british}%
 & \selectlanguage{american}%
A\selectlanguage{british}%
 & \selectlanguage{american}%
\selectlanguage{british}%
 & \selectlanguage{american}%
\selectlanguage{british}%
\tabularnewline
\selectlanguage{english}%
23\selectlanguage{british}%
 & \selectlanguage{english}%
Insulation thickness of external walls {[}mm{]}\selectlanguage{british}%
 & \selectlanguage{english}%
5-100\selectlanguage{british}%
 & \selectlanguage{american}%
A,B1,B2\selectlanguage{british}%
 & \selectlanguage{english}%
90-100\selectlanguage{british}%
 & \selectlanguage{american}%
B3,B4\selectlanguage{british}%
\tabularnewline
\selectlanguage{english}%
24\selectlanguage{british}%
 & \selectlanguage{english}%
Building rotation {[}degrees{]}\selectlanguage{british}%
 & \selectlanguage{english}%
0-180\selectlanguage{british}%
 & \selectlanguage{american}%
A,B1,B2\selectlanguage{british}%
 & \selectlanguage{english}%
0-10\selectlanguage{british}%
 & \selectlanguage{american}%
B3,B4\selectlanguage{british}%
\tabularnewline
\bottomrule
\end{tabular*}}
\par\end{centering}

\selectlanguage{english}%
\caption{List of parameters used in the Morris method first- and second-order
experiments and the different intervals used in the analysis\label{tab:EXP24parameters}. }
\end{sidewaystable}

\par\end{center}

\section*{Figures}

\selectlanguage{american}%
\begin{center}
\begin{figure}[H]
\begin{centering}
\includegraphics[width=0.6\textwidth]{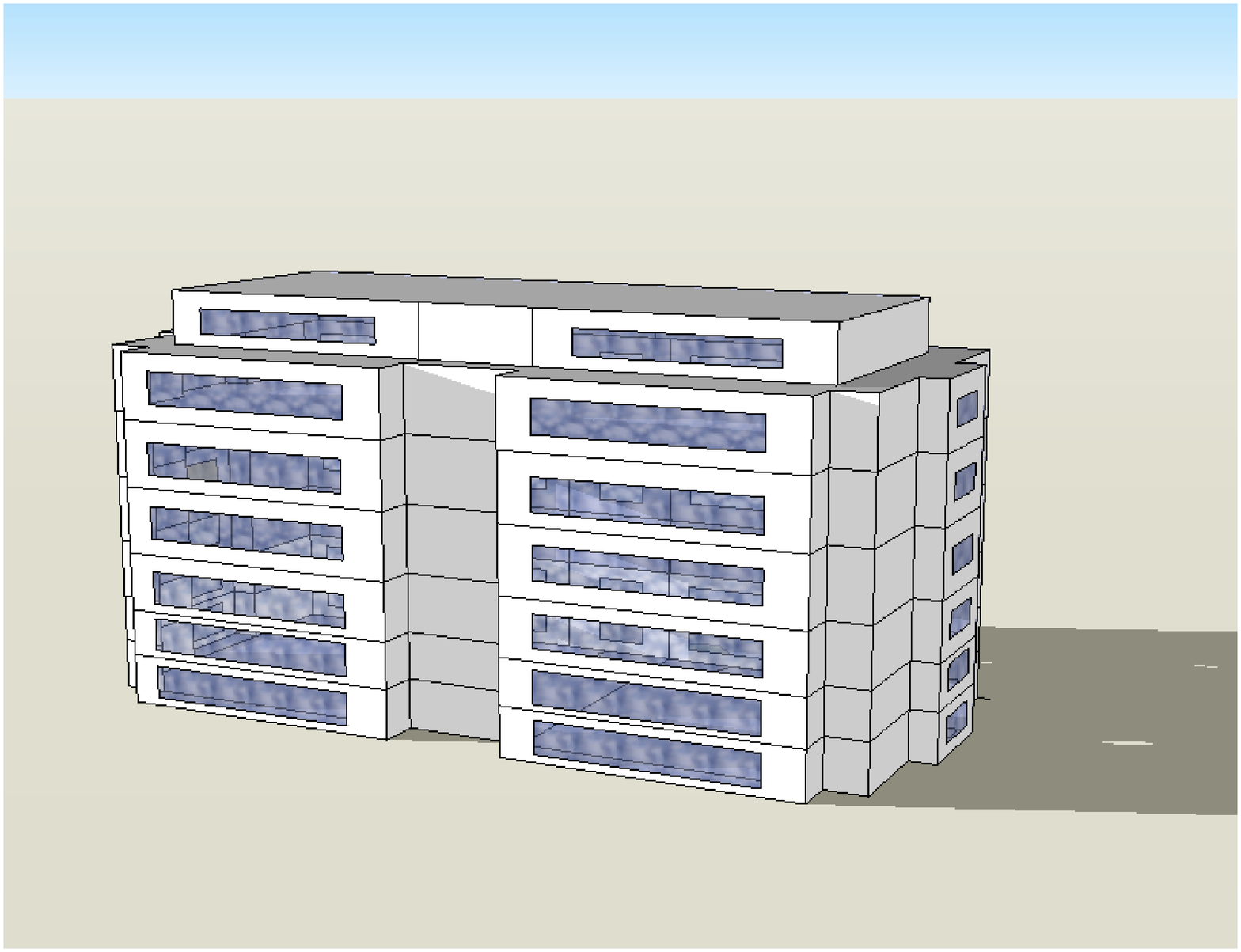}
\par\end{centering}

\caption{\label{fig:The-collective-building}}
\end{figure}

\par\end{center}

\begin{center}
\begin{figure*}[b]
\begin{centering}
\includegraphics[scale=0.5]{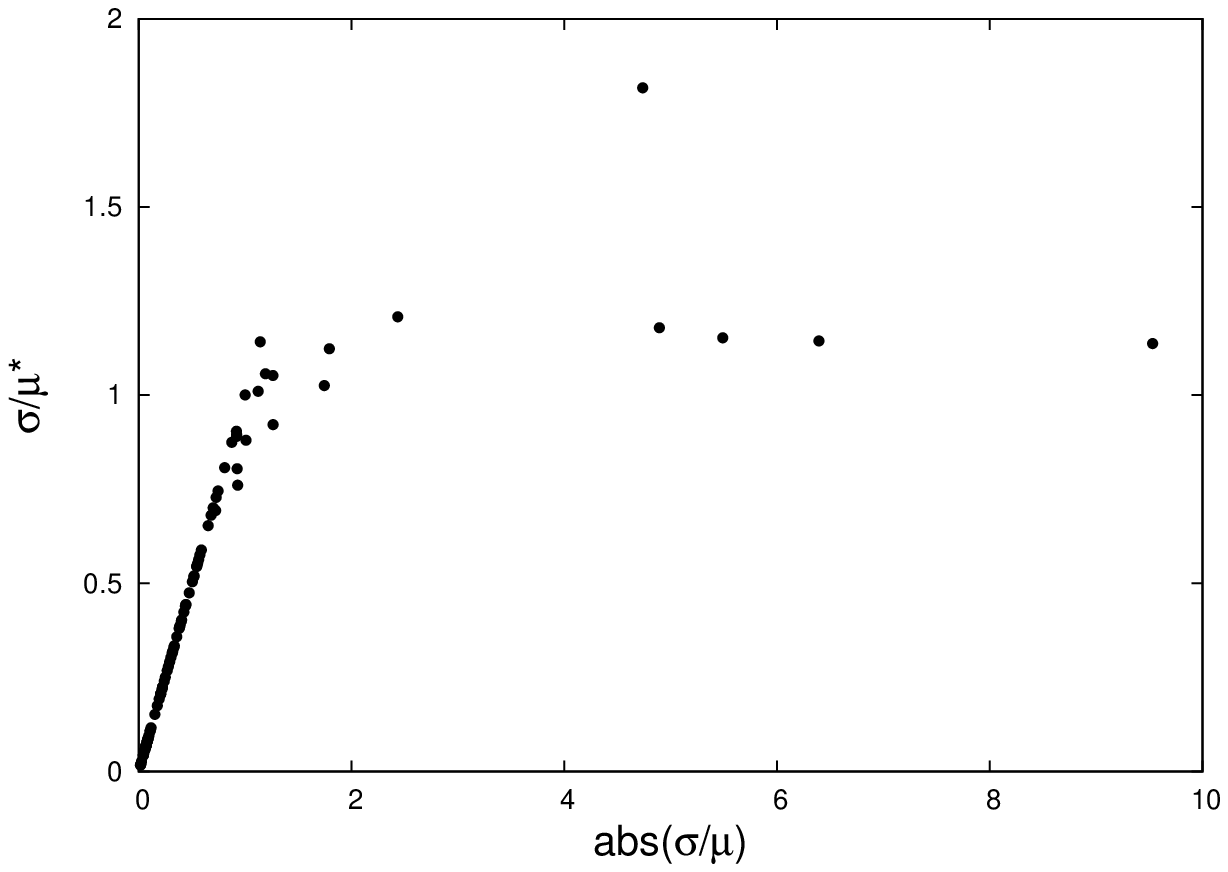}
\par\end{centering}

\caption{\label{fig:Scatterplot}}
\end{figure*}

\par\end{center}

\begin{center}
\begin{sidewaysfigure}
\begin{centering}
\subfloat[Analysis of the elementary effects related to annual heating needs
with a large interval in each parameter and $r=10$.\label{fig:1a}]{\includegraphics[width=0.45\textwidth]{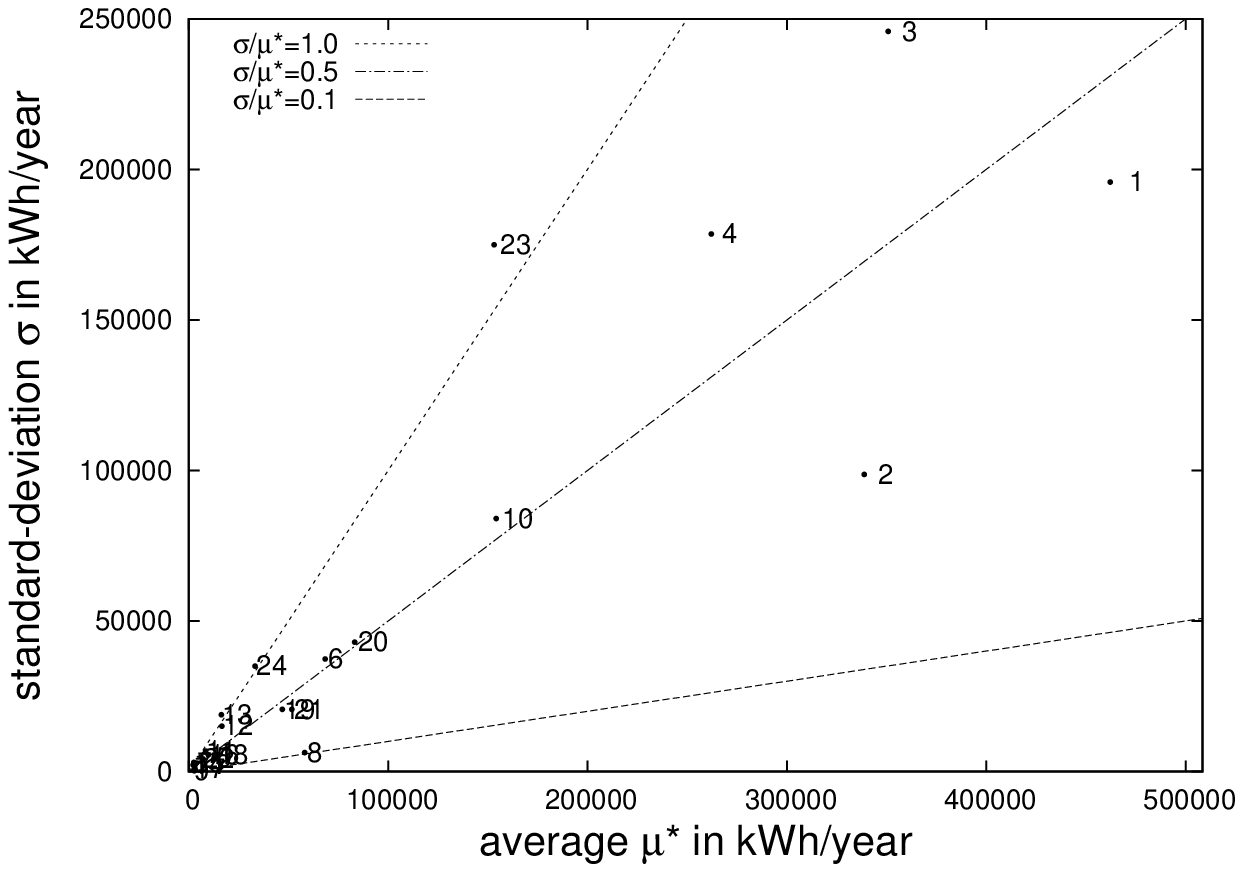}

}\hspace{0.5cm}\subfloat[Analysis of the elementary effects related to the logarithm of annual
heating needs with a large interval in each parameter and $r=10$.
\label{fig:1b}]{\includegraphics[width=0.45\textwidth]{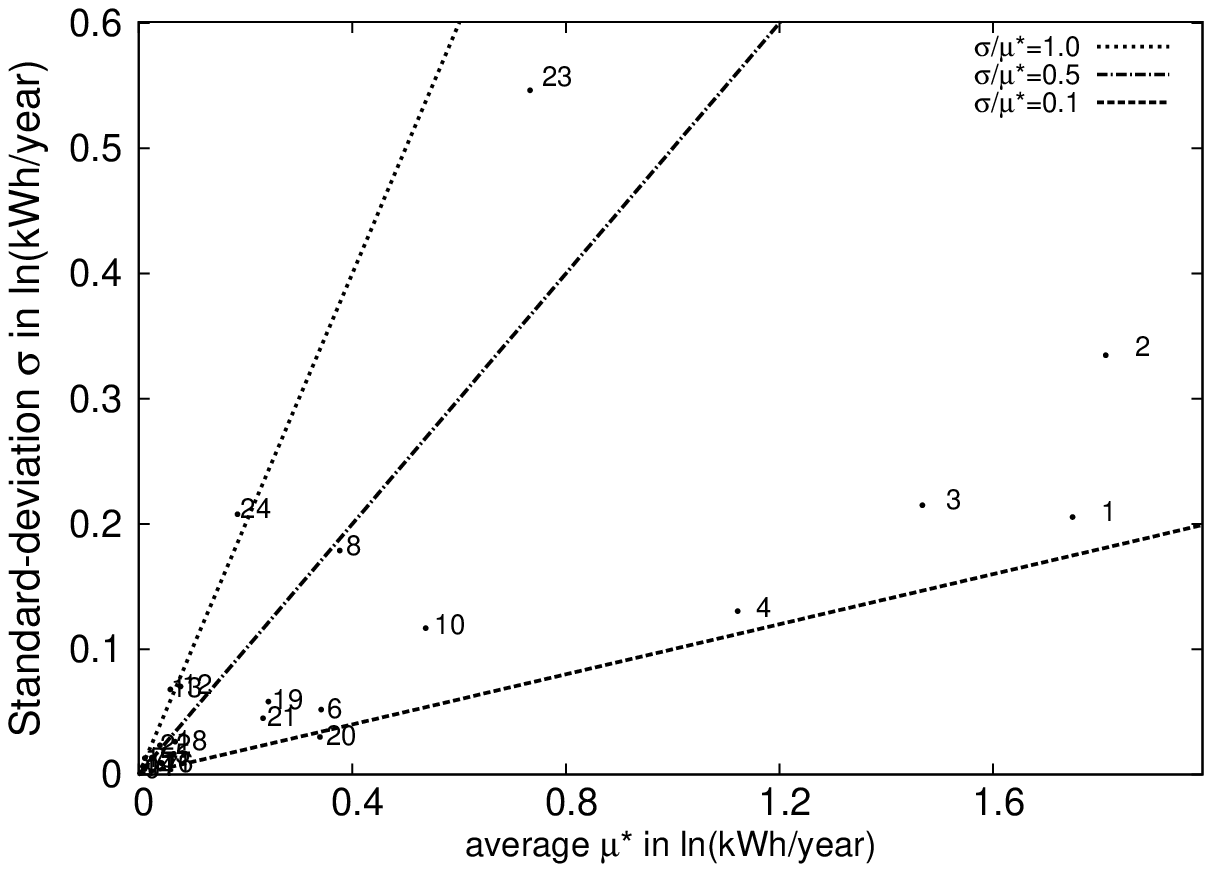}

}
\par\end{centering}

\begin{centering}
\subfloat[ Analysis of the elementary effects related to annual heating needs
per cubic meter with a large interval in each parameter and $r=10$.\label{fig:1c}]{\includegraphics[width=0.45\textwidth]{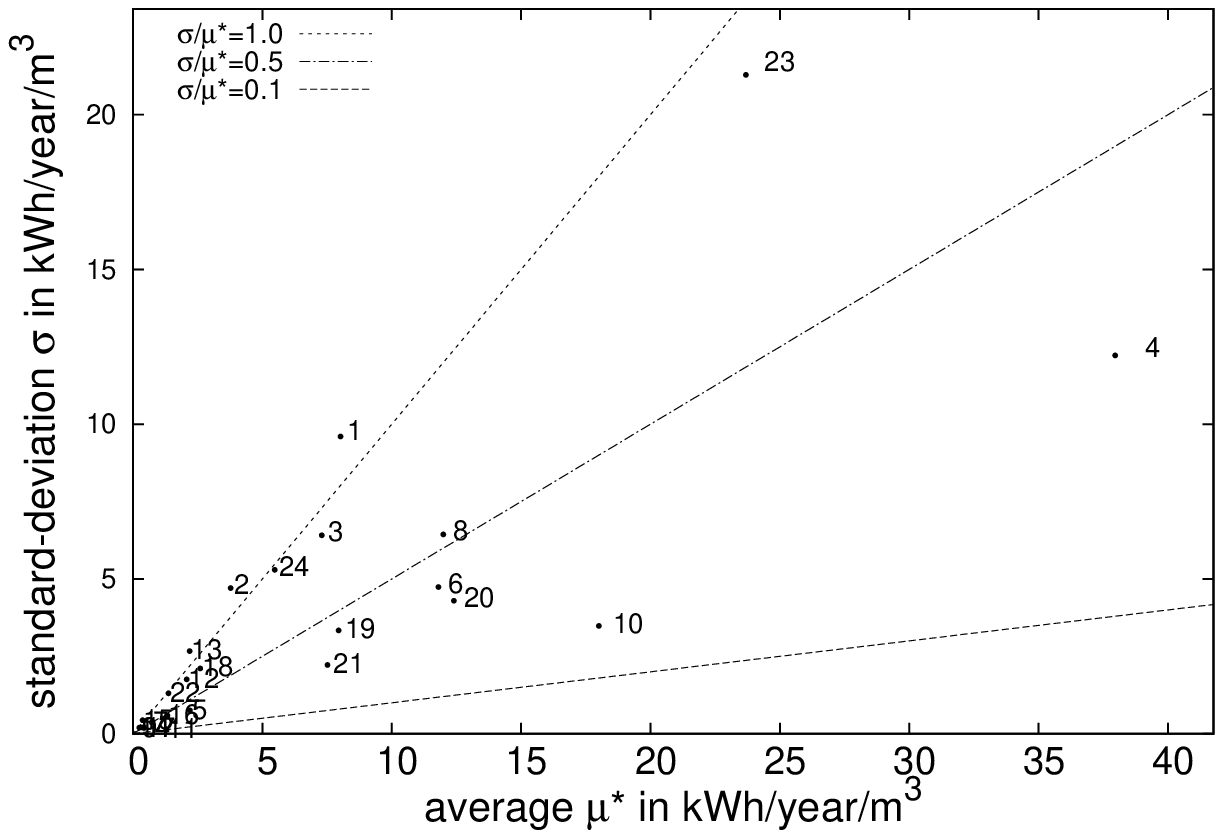}}\hspace{0.5cm}\subfloat[Analysis of the elementary effects related to the natural logarithm
of annual heating needs per cubic meter with a large interval in each
parameter and $r=10$.\label{fig:1d}]{\begin{centering}
\includegraphics[width=0.45\textwidth]{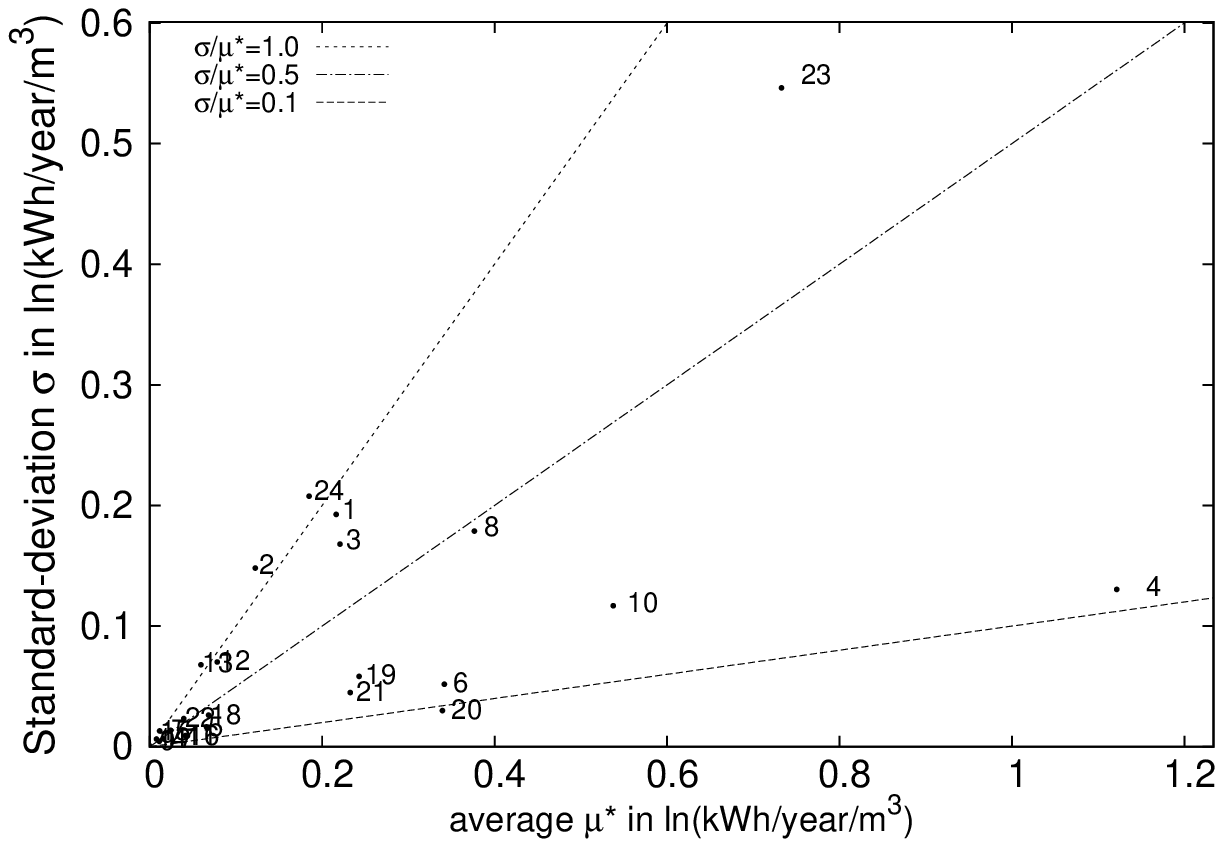}
\par\end{centering}

}
\par\end{centering}

\caption{\label{fig:figure1_experiments}}
\end{sidewaysfigure}

\par\end{center}

\selectlanguage{english}%
\begin{center}
\begin{sidewaysfigure}
\selectlanguage{american}%
\begin{centering}
\subfloat[ Analysis of the elementary effects related to a comfort factor defined
as the mean temperature in summer (July and August) with a large interval
in each parameter and $r=10$\label{fig:2a}]{\begin{centering}
\includegraphics[width=0.45\textwidth]{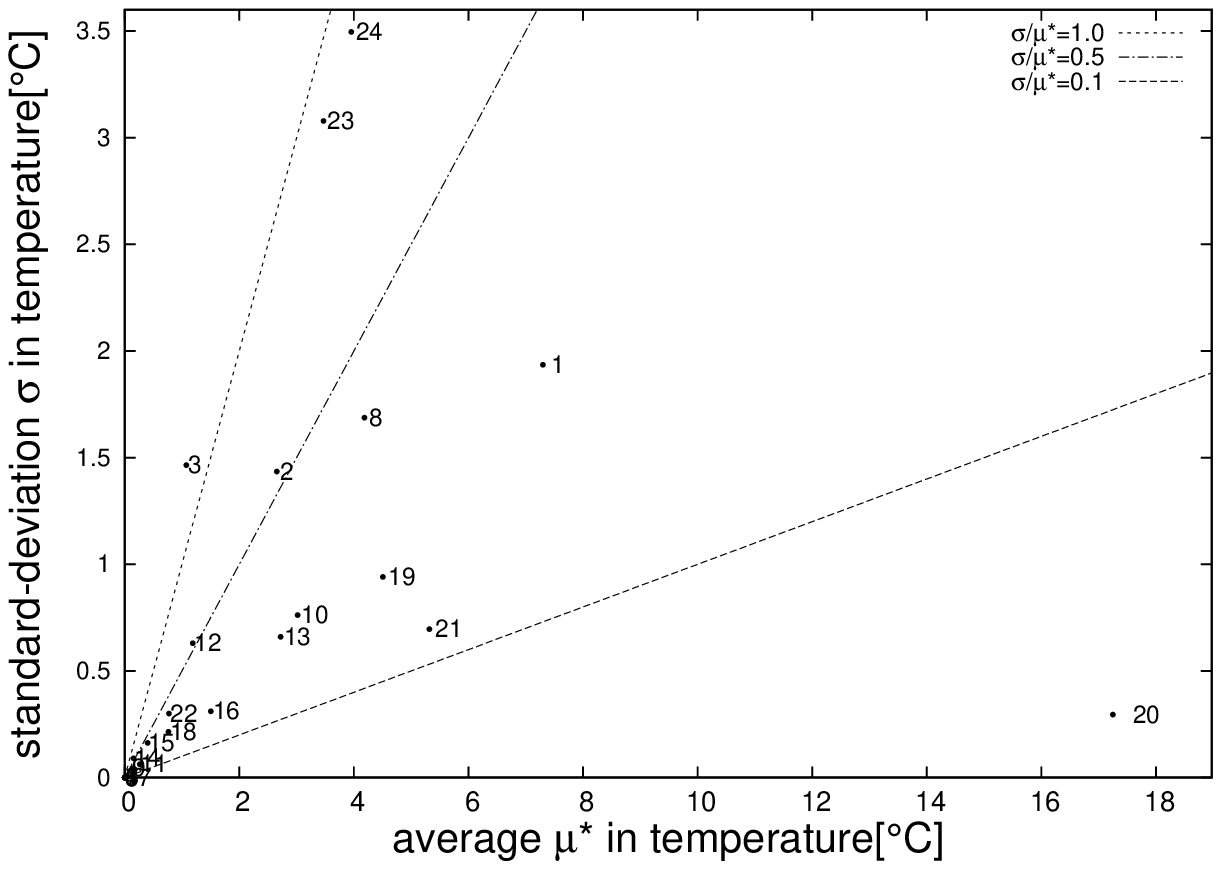}
\par\end{centering}

}\hspace{0.5cm}\subfloat[Analysis of the elementary effects related to power exceeded during
1000 hours/year. Each parameter has a large interval and the analysis
has ten elementary effects per parameter ($r=10$).\label{fig:2c}]{\includegraphics[width=0.45\textwidth]{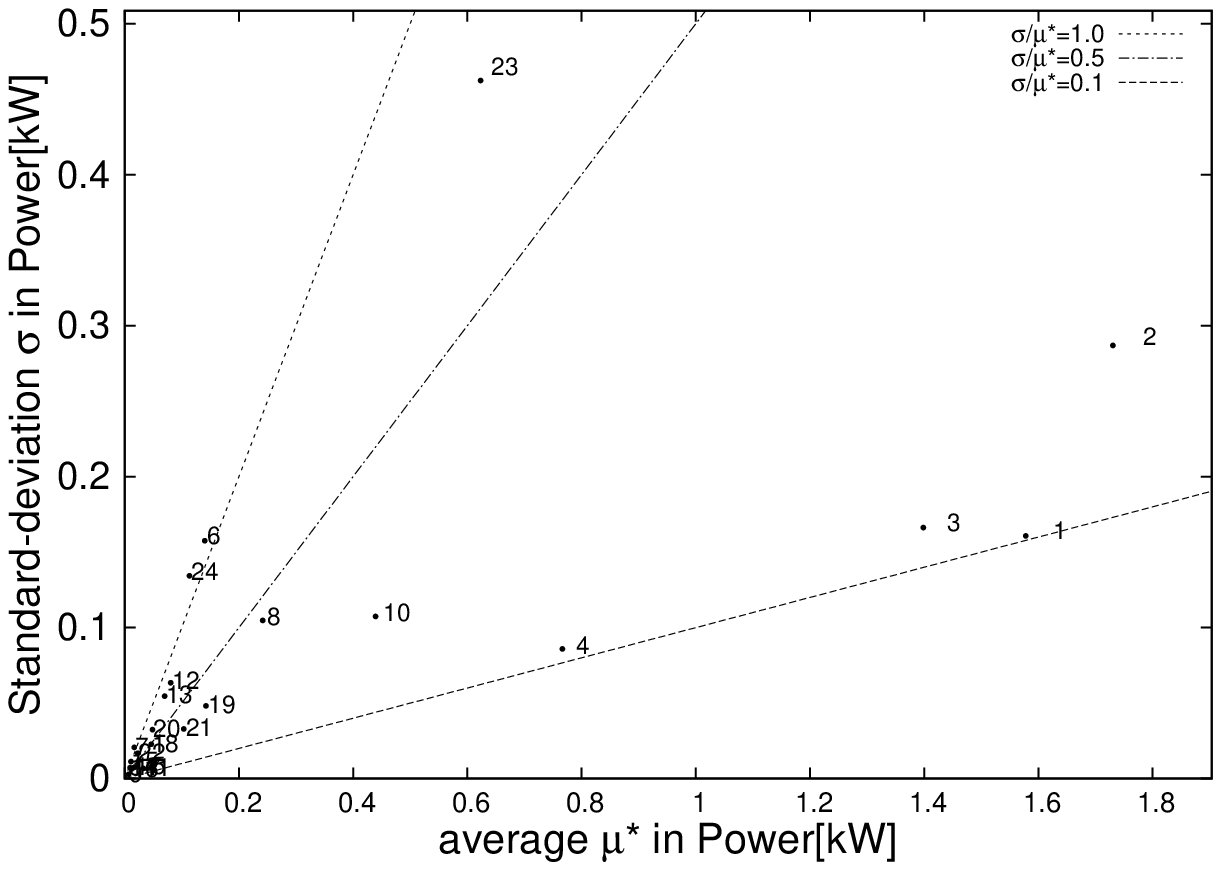}}
\par\end{centering}

\selectlanguage{english}%
\caption{\selectlanguage{american}%
\selectlanguage{english}%
}
\end{sidewaysfigure}

\par\end{center}

\selectlanguage{american}%
\begin{center}
\begin{sidewaysfigure}
\begin{centering}
\subfloat[Analysis of the first-order elementary effects related to annual
heating needs with a large interval in each parameter and $r=10$.
\label{fig:4a}]{\includegraphics[width=0.45\textwidth]{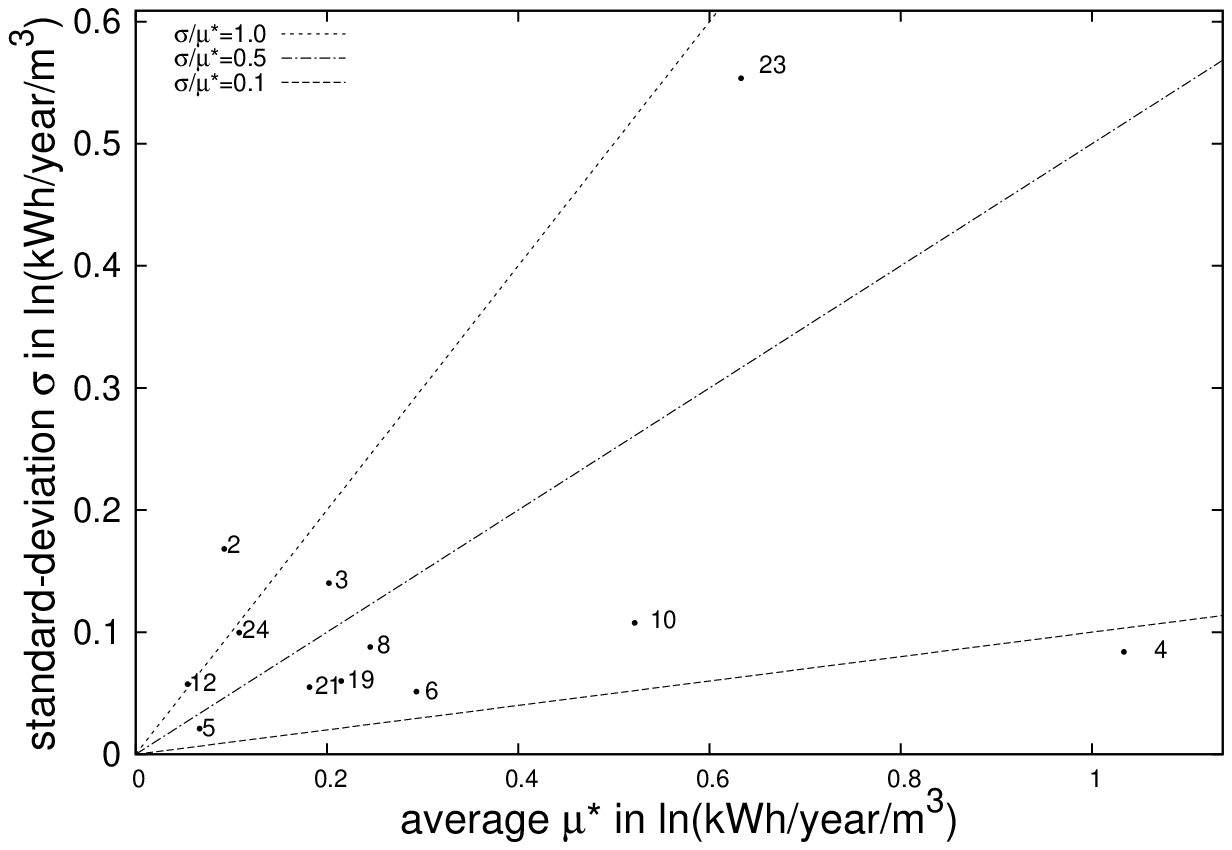}}\hspace{0.5cm}\subfloat[ Analysis of the first-order elementary effects related to annual
heating needs with a large interval in each parameter and $r=10$.
\label{fig:4b}]{\includegraphics[width=0.45\textwidth]{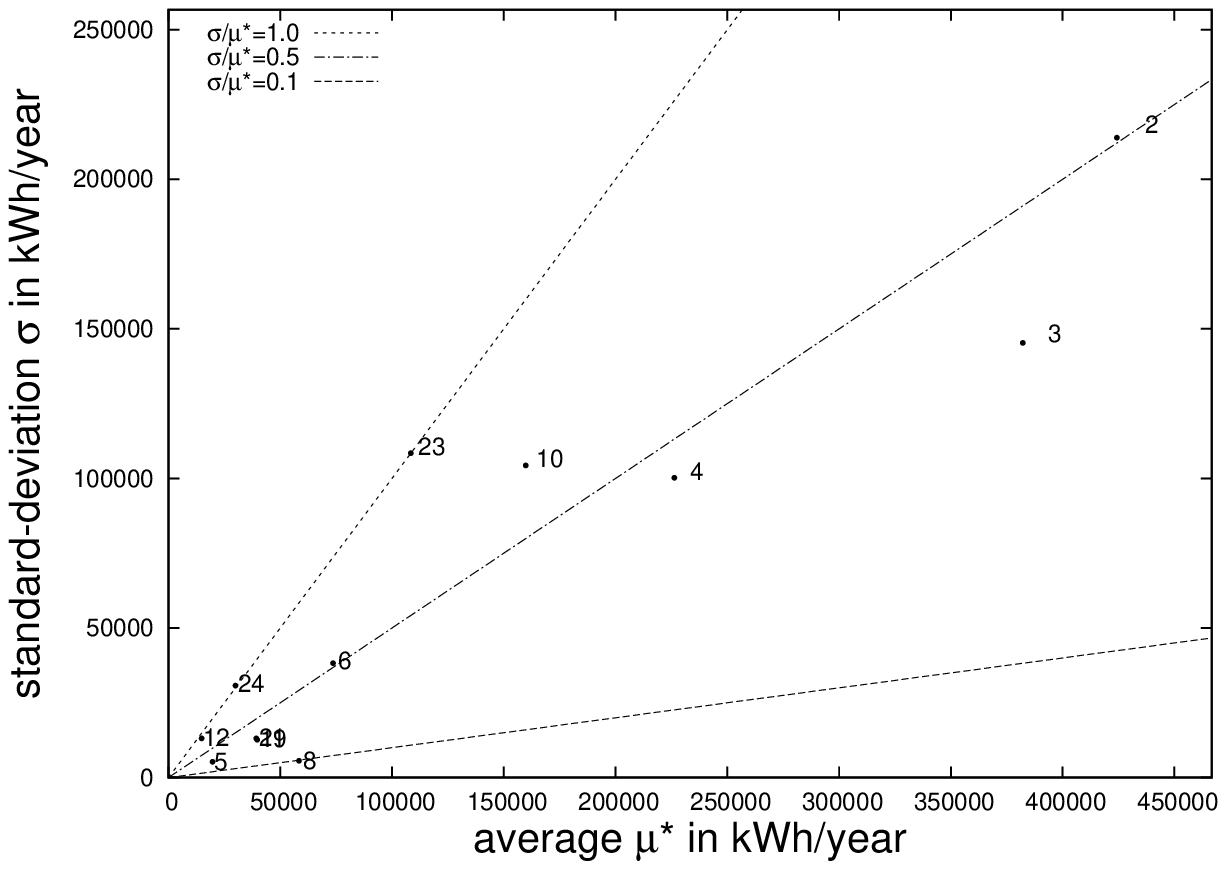}}
\par\end{centering}

\begin{centering}
\subfloat[Analysis of the second-order elementary effects related to annual
heating needs with a large interval in each parameter and $r=10$.
\label{fig:4c}]{\includegraphics[width=0.45\textwidth]{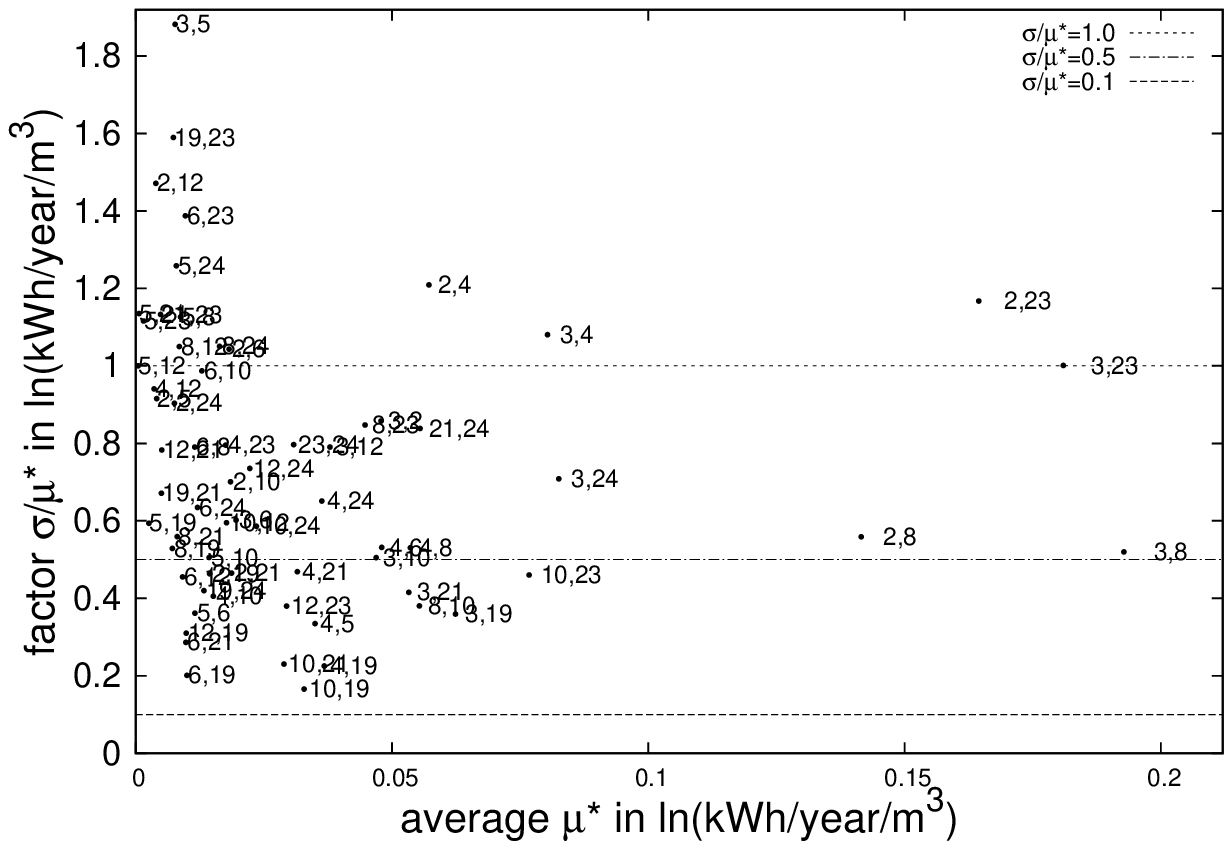}

}\hspace{0.5cm}\subfloat[Analysis of the second-order elementary effects related to annual
heating needs with a large interval in each parameter and $r=10$.
\label{fig:4d}]{\includegraphics[width=0.45\textwidth]{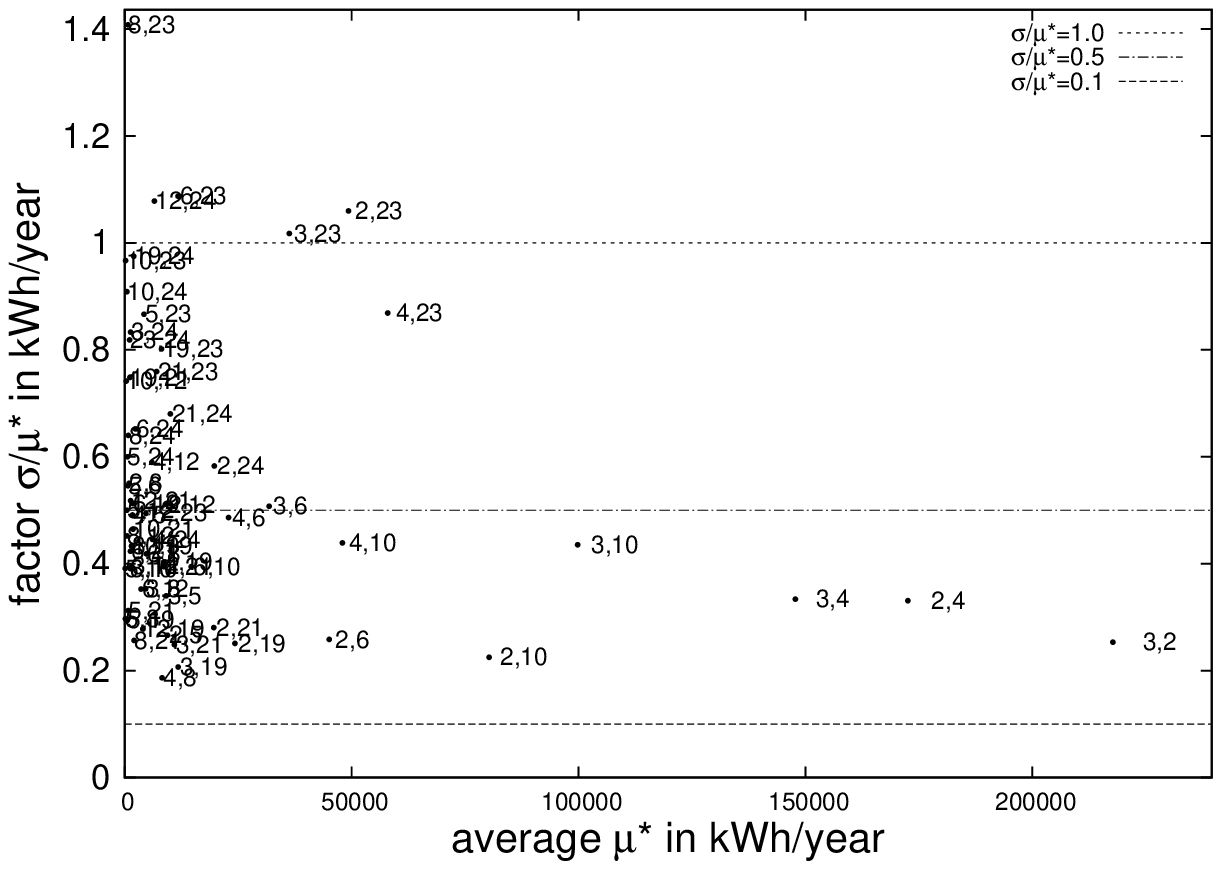}

}
\par\end{centering}

\caption{\label{fig:FIGURE4}}
\end{sidewaysfigure}

\par\end{center}

\begin{center}
\begin{sidewaysfigure}
\begin{centering}
\subfloat[Analysis of the first-order elementary effects related to annual
heating needs with a small interval in each parameter and $r=10$.
\label{fig:5a}]{\includegraphics[width=0.45\textwidth]{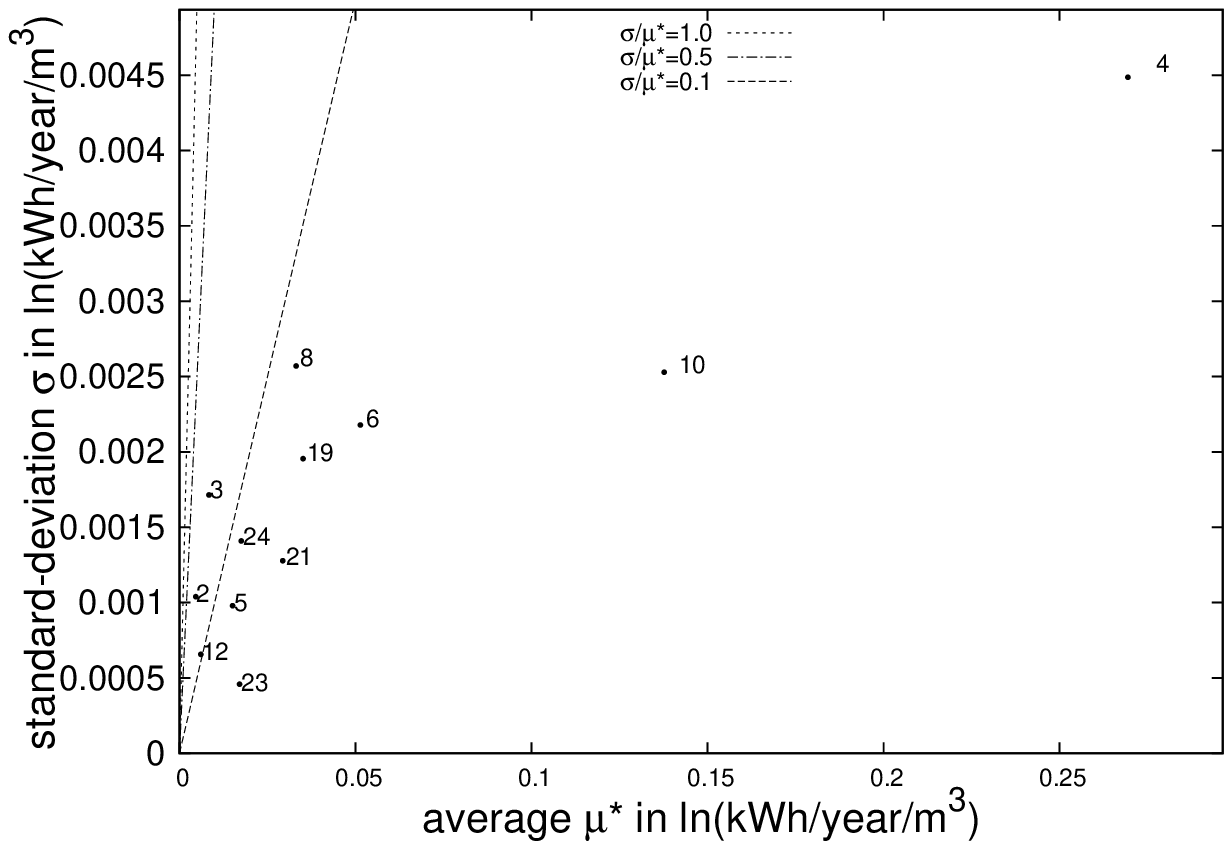}

}\hspace{0.5cm}\subfloat[Analysis of the first-order elementary effects related to annual
heating needs with a small interval in each parameter and $r=10$.
\label{fig:5b}]{\includegraphics[width=0.45\textwidth]{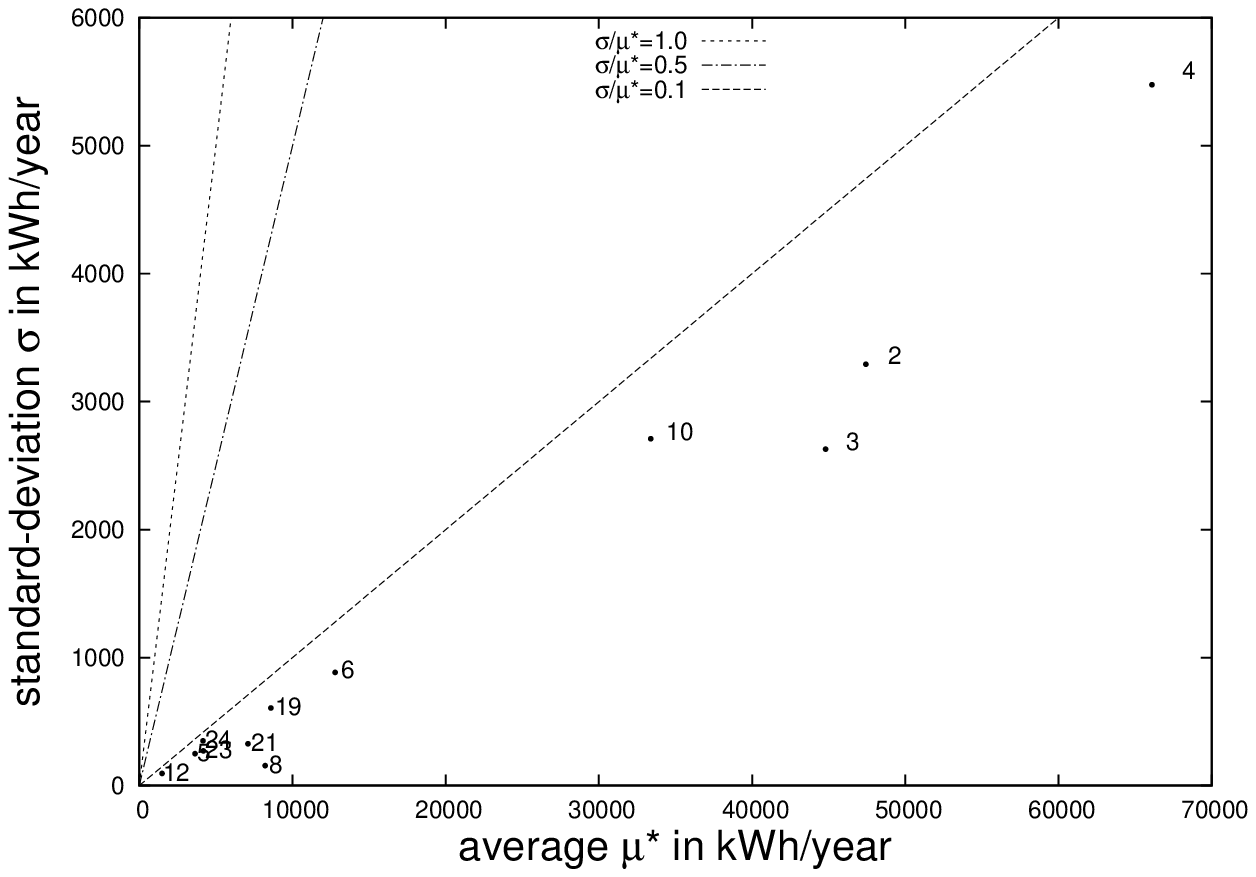}

}
\par\end{centering}

\begin{centering}
\subfloat[Analysis of the second-order elementary effects related to annual
heating needs with a small interval in each parameter and $r=10$.
\label{fig:5c}]{\includegraphics[width=0.45\textwidth]{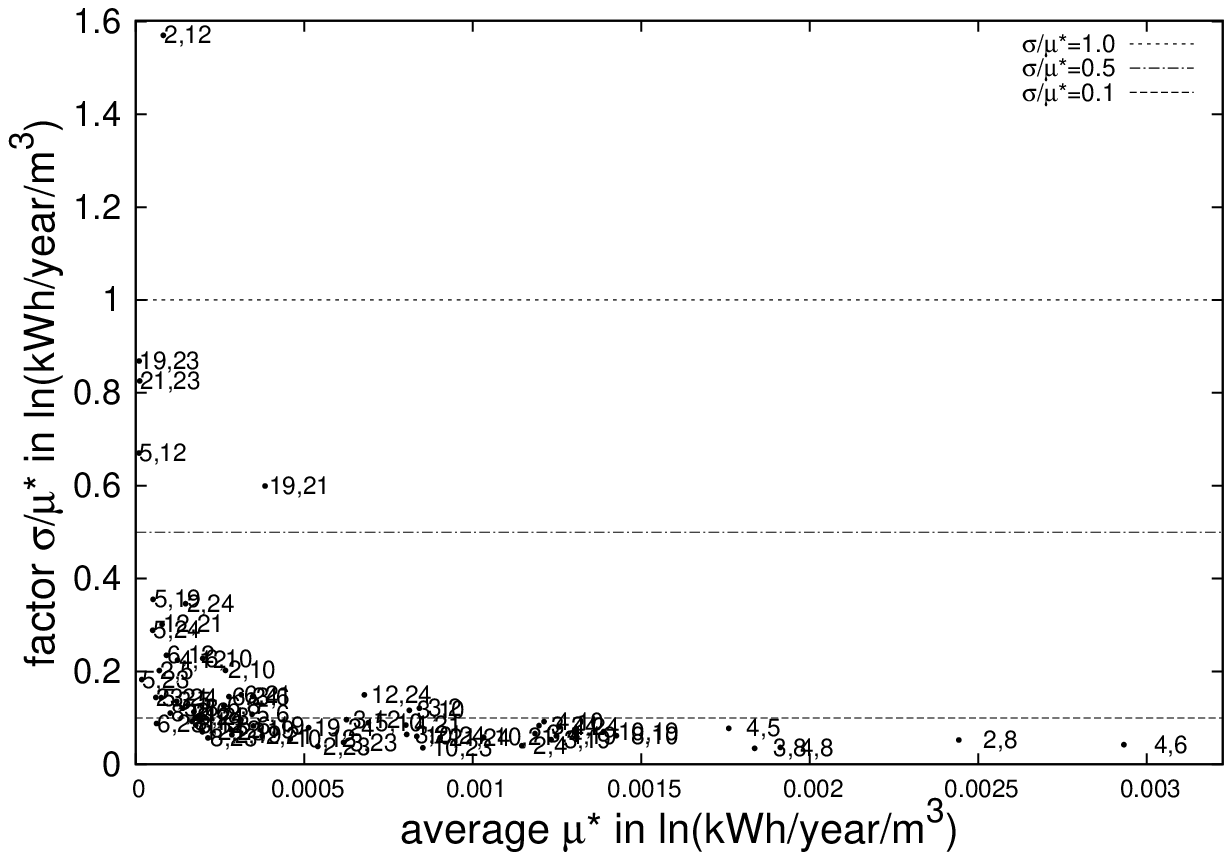}}\hspace{0.5cm}\subfloat[ Analysis of the second-order elementary effects related to annual
heating needs with a small interval in each parameter and $r=10$.
\label{fig:5d}]{\includegraphics[width=0.45\textwidth]{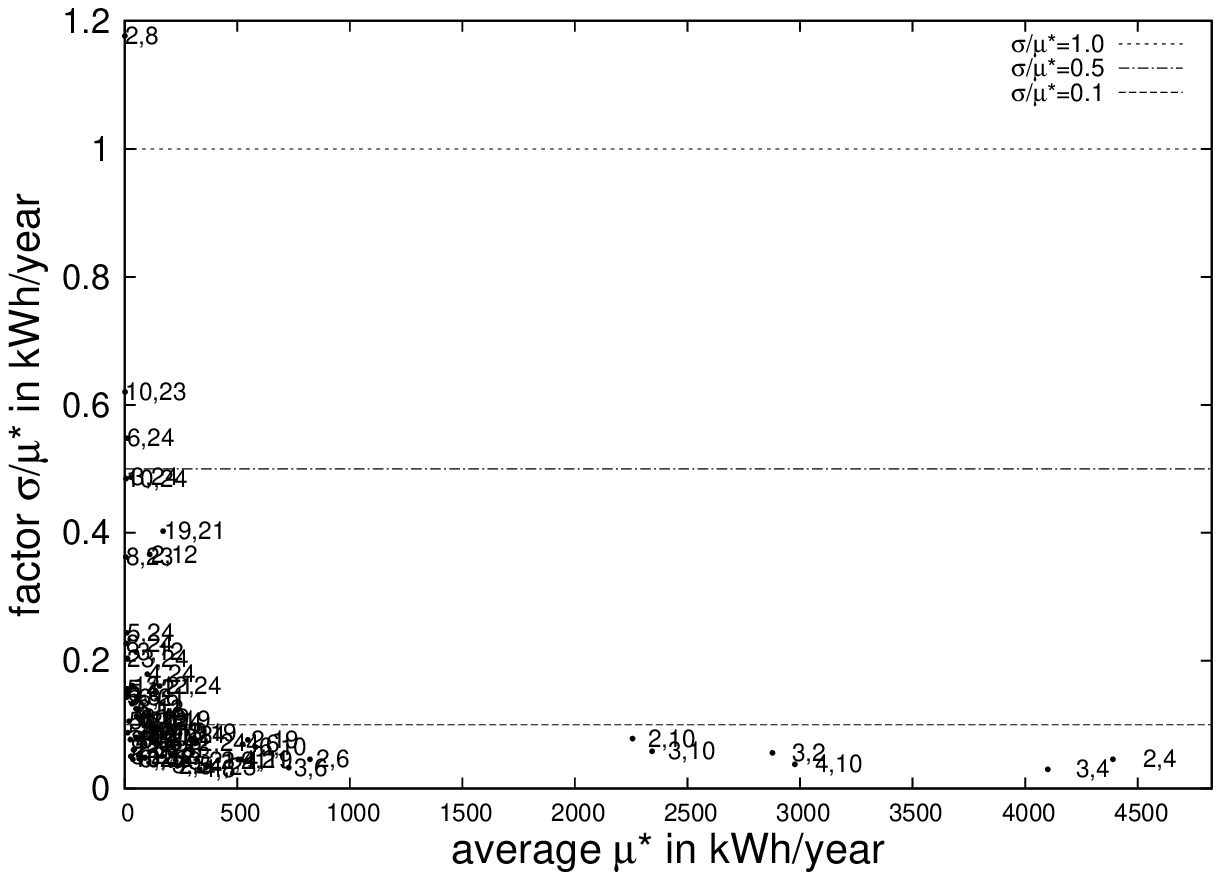}}
\par\end{centering}

\caption{\label{fig:FIGURE5}}
\end{sidewaysfigure}

\par\end{center}
\end{document}